\documentclass[aps, prd, 10pt, notitlepage, superscriptaddress, nofootinbib,numbers,twocolumn]{revtex4-2}

\usepackage[utf8]{inputenc}
\usepackage[T1]{fontenc}
\usepackage{anyfontsize}

\usepackage{amsmath}
\usepackage{amssymb}
\usepackage{amsfonts}
\usepackage{mathrsfs}

\usepackage{microtype}

\usepackage{lipsum,graphicx,afterpage}
\usepackage{epsfig}

\usepackage{bm}

\usepackage{dcolumn}

\usepackage{longtable}
\usepackage{}

\usepackage[dvipsnames]{xcolor}
\usepackage{hyperref}
\hypersetup{
    colorlinks=true,
    citecolor=Purple,
    linkcolor=Purple,
    urlcolor=Purple,
    linktocpage=true,
    breaklinks=true
}

\usepackage{orcidlink}

\usepackage[capitalize]{cleveref}

\begin{document}

\title{Tidal Stretching and Compression in Black Bounce Backgrounds}
\author{T. M. Crispim}
\email{tiago.crispim@fisica.ufc.br}
\affiliation{Departamento de F\'isica, Universidade Federal do Cear\'a, Caixa Postal 6030, Campus do Pici, 60455-760 Fortaleza, Cear\'a, Brazil.}
\author{Marcos V. de S. Silva}
\email{marcosvinicius@fisica.ufc.br}
\affiliation{Departamento de F\'isica, Universidade Federal do Cear\'a, Caixa Postal 6030, Campus do Pici, 60455-760 Fortaleza, Cear\'a, Brazil.}

\author{G. Alencar}
\email{geova@fisica.ufc.br}
\affiliation{Departamento de F\'isica, Universidade Federal do Cear\'a, Caixa Postal 6030, Campus do Pici, 60455-760 Fortaleza, Cear\'a, Brazil.}
\author{Diego S\'aez-Chill\'on G\'omez}
\email{diego.saez@uva.es} 
\affiliation{Department of Theoretical Physics, Atomic and Optics, and Laboratory for Disruptive Interdisciplinary Science (LaDIS), Campus Miguel Delibes, \\ University of Valladolid UVA, Paseo Bel\'en, 7,
47011 - Valladolid, Spain}
\affiliation{Departamento de F\'isica, Universidade Federal do Cear\'a, Caixa Postal 6030,\\ Campus do Pici, 60455-760 Fortaleza, Cear\'a, Brazil.}
\date{\today}

\begin{abstract}
Black bounces are compact objects that combine the structures of regular black holes with those of wormholes. These spacetimes exhibit a rich causal structure and can differ fundamentally from usual black holes. In this work, we study the behavior of the tidal forces by considering different black bounce models. To this end, we start with the geodesic deviation equation and the tidal tensor, from which we compute the radial and angular components of the tidal forces. We find that these components are finite throughout the entire spacetime, including at the wormhole throats. Through the components of the displacement vector, we observe that, unlike the Schwarzschild case, a compression effect on bodies may occur in certain regions.
\end{abstract}


\maketitle

\section{Introduction}
Black holes (BHs) are astrophysical objects that initially emerged as theoretical predictions of general relativity (GR), arising from solutions of the Einstein equations. These objects are notable for their causal structure, possessing a surface of no return known as the event horizon. BHs are the simplest solutions of the Einstein equations, as they are essentially described by three parameters: mass, charge, and angular momentum \cite{Ruffini:1971bza,Robinson:2004zz}. However, due to discharge processes \cite{Gibbons:1975kk,Goldreich:1969sb,Blandford:1977ds}, it is expected that astrophysical BHs do not possess charge and therefore have only mass and rotation \cite{Kerr:1963ud}. It is important to emphasize that, in the context of minicharged dark matter, the discharge process does not occur in the same way, which allows the existence of charged solutions \cite{Cardoso:2016olt}.

Although not expected astrophysically, charged BHs, such as the Reissner-Nordström solution, are of great theoretical interest, even in static cases. These models can exhibit important features such as a Cauchy horizon, zero Hawking temperature, naked singularities, superradiance, and equality between the scattering and absorption coefficients for electromagnetic and gravitational waves \cite{Pugliese:2010ps,Pugliese:2011py,Ghosh:1994mm,Pavon:1991kh,Benone:2015bst,Sanchis-Gual:2015lje,Oliveira:2011zz,Crispino:2015gua}. Some of these phenomena may also occur in uncharged cases when rotating solutions are considered.

A problem that arises in the context of BHs is the presence of singularities \cite{Penrose:1964wq}. In many cases, the physical quantities of interest diverge in these regions \cite{Gubser:2000nd}. However, singularities can be defined as points where geodesics are incomplete \cite{Bronnikov:2012wsj}. Although they usually do not represent a phenomenological issue due to being hidden behind an event horizon \cite{Penrose:1969pc}, singularities may indicate a breakdown of classical theories in describing physics in certain regions.

In GR, one way to address the problem of singularities is through the study of regular BHs, that is, BHs without singularities in their interior \cite{Ansoldi:2008jw}. The first regular solution was proposed by Bardeen as a generic example of a regular metric and was later derived from Einstein equations by Beato and Garcia, who showed that the Bardeen metric can be interpreted as a magnetically charged BH sourced by a nonlinear electrodynamics \cite{Ayon-Beato:2000mjt,Rodrigues:2018bdc}. The center of these solutions features a de Sitter or Minkowski like core instead of a singularity \cite{Bronnikov:2024izh,Simpson:2019mud,Culetu:2014lca,Culetu:2015cna}. The study of regular BHs supported by nonlinear electrodynamics has been extensively developed in the literature, since the electrodynamics itself modifies important properties of BHs \cite{Bronnikov:2022ofk,Bronnikov:2000vy,Ayon-Beato:1998hmi,Ayon-Beato:1999kuh,Ayon-Beato:1999qin,Dymnikova:2004zc,Balart:2014cga,Kruglov:2017fck,deSousaSilva:2018kkt,Rodrigues:2017yry,Rodrigues:2019xrc,Junior:2020zdt}. One of the most important examples is that the trajectory of the photons is influenced by the presence of nonlinear electrodynamics \cite{Habibina:2020msd,Toshmatov:2021fgm,dePaula:2024yzy}, which alters the shadow of these BHs \cite{Kruglov:2020tes,Allahyari:2019jqz,Stuchlik:2019uvf,dePaula:2023ozi}. Another significant feature is how the thermodynamics of these BHs is modified by the first law of thermodynamics \cite{Maluf:2018lyu,Zhang:2016ilt,Ma:2014qma,Kumar:2020cve,Rodrigues:2022qdp}. Since these are charged solutions, there are also systems that exhibit superradiance \cite{dePaula:2024xnd,Dolan:2024qqr,dePaula:2025kif}.

Another way to address the problem of singularities in solutions is through spacetimes known as black bounces (BBs). In this type of spacetime, the singularity and its surroundings are removed and replaced by the throat of a wormhole \cite{Bolokhov:2024sdy}. The first BB model was proposed by Hochberg and Visser in 1997 and was later further studied by Simpson and Visser in 2018 \cite{Visser:1997yn,Simpson:2018tsi}. This metric, known as the Simpson-Visser (SV) spacetime, can be obtained through a regularization process of the Schwarzschild metric by introducing the parameter $a$, which regularizes the spacetime by replacing $x \rightarrow \sqrt{x^2 + a^2}$ with the metric coefficients. Depending on the value of this parameter, the solution can represent a singular BH, a regular BH, or a wormhole. From the perspective of BH shadows, these solutions can closely mimic their singular counterparts \cite{Guerrero:2021ues,Lima:2021las,Vagnozzi:2022moj,Tsukamoto:2020bjm}. Several BB models are constructed using this regularization method \cite{Simpson:2019cer,Franzin:2021vnj,Furtado:2022tnb,Lima:2022pvc,Lima:2023jtl,Crispim:2024nou,Crispim:2024yjz} (for general aspects and consistency, see Ref. \cite{Alencar:2025jvl}); however, other models arise from different approaches \cite{Huang:2019arj,Lobo:2020ffi,Fitkevich:2022ior,Rodrigues:2022mdm,Rodrigues:2022rfj,Pereira:2023lck,Rodrigues:2023fps,Pereira:2024gsl}.

These spacetimes, as they are mostly proposed, are not solutions of the field equations of known theories. However, it is possible to perform a reverse engineering procedure, where the spacetimes are imposed as solutions of the field equations, and from this the type of source that generates such solutions is determined \cite{Bronnikov:2022bud}. The most common type involves a phantom scalar field, which may also be of the $k$-essence type, combined with a nonlinear electrodynamics. This type of source has been found for several types of solution, such as spherically symmetric configurations \cite{Bronnikov:2021uta,Canate:2022gpy,Rodrigues:2023vtm,Pereira:2024rtv,Rodrigues:2025plw}, cylindrically symmetric solutions \cite{Bronnikov:2023aya,Lima:2023arg}, $2+1$ dimensional models \cite{Alencar:2024yvh}, in modified gravity theories \cite{Junior:2024vrv,Alencar:2024nxi,Silva:2025fqj}, and even in scenarios with Lorentz symmetry breaking \cite{Pereira:2025xnw}.

However, it is expected that once we achieve a complete quantum description of gravity, many of the current problems of GR will disappear, including singularities. In this context, one possible candidate theory is loop quantum gravity (LQG), which aims to quantize spacetime itself \cite{Ashtekar:2021kfp}. Inspired by this idea, it is possible to take an effective approach in order to introduce quantum corrections into gravity. In \cite{Alonso-Bardaji:2021yls,Alonso-Bardaji:2022ear}, holonomy corrections were considered in the simplest non-homogeneous spacetime, the spherically symmetric one, thus obtaining a singularity-free BH solution containing a minimal spacelike hypersurface inside the horizon. In this way, we see that BB solutions can also naturally emerge from effective models inspired by LQG \cite{Peltola:2008pa,Bodendorfer:2019cyv,Zhang:2024khj,Zhang:2024ney}.

Tidal forces are the result of the variation of the gravitational field of a massive body along the extent of another body \cite{Hobson:2006se}. This difference in the experienced forces can cause deformations in the affected body, typically stretching in one direction and compression in another. If the gravitational field is sufficiently intense, an astrophysically viable scenario, for example, when stars are located near BHs \cite{Gezari:2021bmb}, there will be a point at which the tidal forces exceed the internal forces of the secondary body, leading to its rupture, tidal disruption \cite{Rees:1988bf,Evans:1989qe}. This point is called the Roche radius \cite{Aggarwal:1974}. In the case of a Schwarzschild BH, an infalling body experiences tidal forces that stretch it in the radial direction and compress it in the tangential directions \cite{Hobson:2006se}. These forces increase as the body approaches the BH, diverging at the center, the singularity. Depending on the mass of the BH, the Roche radius may lie inside the event horizon such that the tidal disruption is not observable \cite{Hobson:2006se}.

Different BH solutions can exhibit different tidal forces and, consequently, different properties. In the case of the Reissner-Nordström BH, the radial and tangential components of the tidal force reach a maximum/minimum and then change sign, allowing a compression effect on the infalling body instead of a stretching one \cite{Crispino:2016pnv,Sharif:2018gzj}. The components of the tidal force may also diverge in this case, but the test particles falling into this spacetime reach a point of closest approach where the tidal force remains finite \cite{Crispino:2016pnv}. For regular BHs, a sign change in the tidal force components is also observed; however, unlike the Reissner-Nordström case, there are no divergences in these components \cite{Sharif:2018gaj,Lima:2020wcb}. On the other hand, there are solutions in which the tidal forces diverge even at the event horizon \cite{Nandi:2000gt}. Therefore, we see that, depending on the BH model considered, the tidal forces may display quite distinct behaviors \cite{LimaJunior:2022gko,Masi:2007mn,Shahzad:2017vwi,LimaJunior:2020fhs,Andre:2020qey,Vandeev:2021yan,Uniyal:2022ouc,Arora:2023ltv,Toshmatov:2023anz,Albacete:2024qja}, and such modifications can directly affect the Roche radius.

This work is organized as follows. In Section \ref{SEC:spacetime}, we present the spacetime for the four BB models, disusing general properties such as event horizons and throats. In Section \ref{SEC:geodesics}, we studied the geodesics of massive, uncharged particles falling radially into a BH. In Section \ref{SEC:TF}, we developed the formalism required for the study of tidal forces, such as the geodesic deviation equation and the tidal tensor, and applied this formalism to the four models presented in Section \ref{SEC:spacetime}.  Finally, in Section \ref{SEC:conclusions}, we present our conclusions of our work and discuss future research directions.

\section{spacetimes}\label{SEC:spacetime}
In the coordinate system where the radial coordinate can take values between $-\infty < x < + \infty$ (including zero), BB geometries generally have their line element written in the form of
\begin{eqnarray}
    ds^2 = -f(x)dt^2 + \frac{dx^2}{g(x)} + \Sigma(x)^2d\Omega_2^2,\label{metricageral_x}
\end{eqnarray}
where $d\Omega_2^2=d\theta^2+\sin^2\theta d\phi^2$.

However, for reasons that will become clear shortly, it is always possible to perform the coordinate transformation given by $r^2 = \Sigma(x)^2$, where now the radial coordinate takes a minimum value given by $r=\Sigma(x = 0)$. With this, we can rewrite the line element in a more convenient form, where we will call the radial coordinate simply $r$. Thus, we have
\begin{equation}
  \label{metricageral}  ds^2 = -f(r)dt^2 + \frac{dr^2}{h(r)} + r^2d\Omega_2^2,
\end{equation}
where it is important to remember that the radial coordinate cannot take the value of $r = 0$. That is, it has a minimum value, different from zero, which will depend on the specific BB geometry.

Next, we will discuss the general aspects of the BB geometries that will be addressed in this paper.

\subsection{SV spacetime}
The SV spacetime is the simplest BB \cite{Simpson:2018tsi,Visser:1997yn}. This spacetime can be obtained through a regularization process applied to the Schwarzschild metric coefficients by making the substitution $x \to \sqrt{x^2 + a^2}$. Therefore, the SV metric is a special case of the line element \eqref{metricageral_x}, with:
\begin{equation}
    f(x)=g(x)=1-\frac{2m}{\sqrt{x^2+a^2}},\quad \Sigma(x)=\sqrt{x^2+a^2}.
\end{equation}
This is the most well-known and extensively studied form of the SV spacetime. However, we can apply the coordinate transformation discussed earlier and express this solution as a special case of the line element \eqref{metricageral} for \cite{Franzin:2023slm}
\begin{eqnarray}
  \label{SVEq}  f(r) = 1 - \frac{2m}{r}, \quad h(r) = \left(1 - \frac{2m}{r}\right)\left(1 - \frac{a^2}{r^2}\right),
\end{eqnarray}
where the radial coordinate takes values between $a < r < + \infty$.

It is well known in the literature that, depending on the values of the parameter $a$, this line element can describe a regular BH, a one-way traversable wormhole, a two-way traversable wormhole, or a singular BH. In the coordinate system that we are using, the event horizon is located at $r = r_h = 2m$, as is usually the case for Schwarzschild, and the throat radius is located at $r_t = a$. For $a > 2m$, the throat radius becomes larger than the horizon radius, and in this situation the spacetime represents a wormhole that can be traversed in both directions.

\subsection{Bardeen-Type}
\begin{figure}[htb]
    \centering
    \includegraphics[width=1.\linewidth]{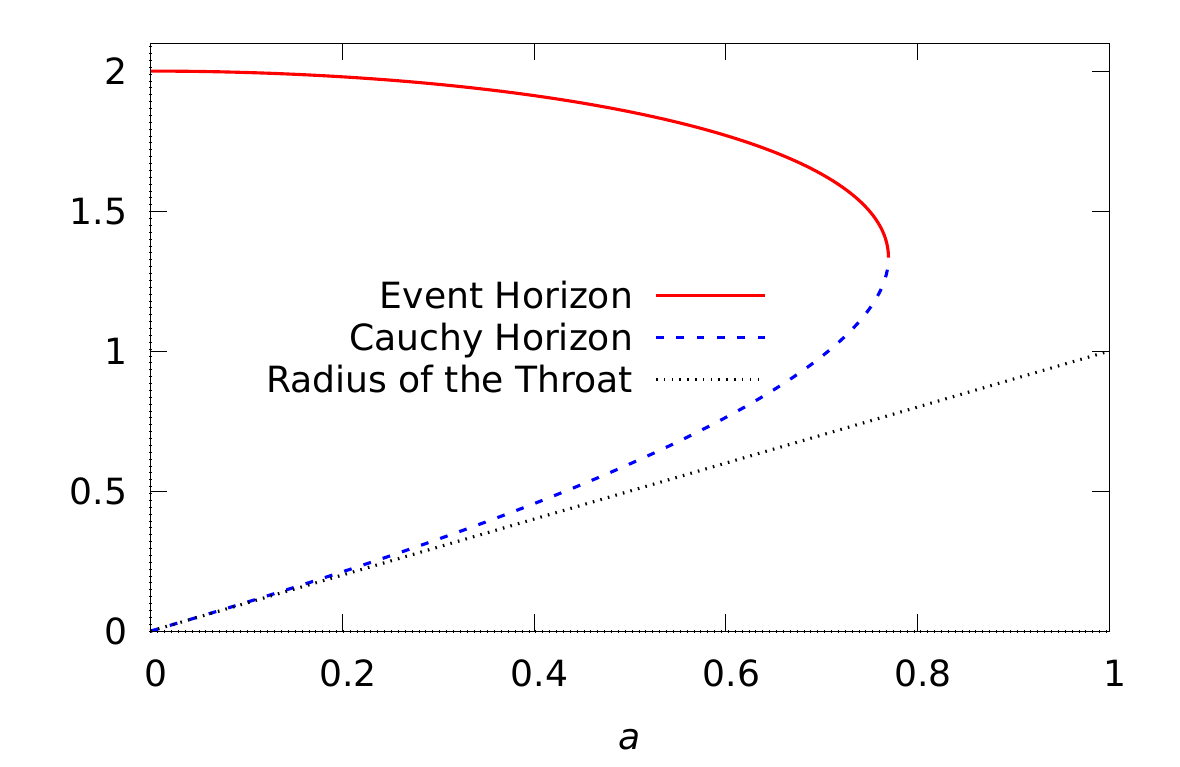}
    \caption{Behavior of the horizon radii and the wormhole throat radius for the Bardeen-type solution as a function of the regularization parameter. We choose $m=1$.}
    \label{fig:raios_bardeen}
\end{figure}

A second BB model widely studied in the literature is the Bardeen-type \cite{Lobo:2020ffi}. In the original coordinate system in which it was initially proposed, this spacetime is described by the line element \eqref{metricageral_x}, whose metric coefficients are
\begin{equation}
    f(x)=g(x)=1-\frac{2mx^2}{\left(x^2+a^2\right)^{3/2}}, \quad \Sigma(x)=\sqrt{x^2+a^2}.
\end{equation}
In the coordinate system we are considering, this solution will be described by the line element \eqref{metricageral}, whose coefficients are
\begin{equation}
 \label{bardeef}   f(r)=1- \frac{2m}{r} + \frac{2ma^2}{r^3},
\end{equation}
\begin{equation}
  \label{bardeenh}  h(r)=\left(1- \frac{2m}{r} + \frac{2ma^2}{r^3}\right)\left(1-\frac{a^2}{r^2}\right).
\end{equation}
This solution features two event horizons and a wormhole throat hidden behind the horizons, an event horizon, and a Cauchy horizon. As in the SV case, the radius of the throat of the wormhole is given by $r_{t} = a$. The analytical expressions for the horizons are complicated, but we can analyze their behavior in terms of the regularization parameter graphically. In Fig. \ref{fig:raios_bardeen}, we observe the behavior of the radii of the event horizon, the Cauchy horizon, and the wormhole throat. There exists an extremal value, $a = a_{ext} = \frac{4m}{3\sqrt{3}}$, above which there are no horizons, resulting in a wormhole configuration. For $a < a_{ext}$, the throat of the wormhole remains hidden behind the horizons. The causal structure of this spacetime has been studied in detail in \cite{Lobo:2020ffi}.

\subsection{Holonomy corrected}

A third BB model we can study is the holonomy corrected Schwarzschild BH. This spacetime is described by the metric \eqref{metricageral}, whose coefficients are given by \cite{Alonso-Bardaji:2021yls,Alonso-Bardaji:2022ear}
\begin{equation}\label{HCEq}  
f(r) = 1 - \frac{2m}{r}, \quad h(r) = \left(1 - \frac{2m}{r}\right)\left(1 - \frac{a}{r}\right),
\end{equation}
where $a<2m$. Similarly to the SV case, this solution features an event horizon located at $r=2m$ and a wormhole throat with radius $r_t=a$. An analysis of the causal structure of this spacetime can be found in \cite{Moreira:2023cxy}.

Unlike the other BB models, this spacetime will not exhibit traversable wormholes since $a<2m$, so that the throat radius cannot be larger than the event horizon radius.

This model can also be written in a different coordinate system, similar to the other models, considering $r^2 = x^2 + a^2$. In this case, we have the line element \eqref{metricageral_x} with
\begin{equation}
    f(x)=1-\frac{2m}{\sqrt{x^2+a^2}},
\end{equation}
\begin{equation}
    g(x)=\left(1-\frac{2m}{\sqrt{x^2+a^2}}\right)\left(1+\frac{a}{\sqrt{x^2+a^2}}\right)^{-1},
\end{equation}
with $\Sigma(x)=\sqrt{x^2+a^2}$. Adopting this coordinate system for the model is not particularly advantageous, as it leads to three functions depending on the radial coordinate $x$, making the problem more difficult from an algebraic point of view.

\subsection{Polymerized BH}
The last BB model we will consider is the polymerized BH. This spacetime is described by the line element \eqref{metricageral}, whose coefficients are given by \cite{Peltola:2008pa}:
\begin{equation}\label{Poly_f}
    f(r)=\sqrt{1-\frac{a^2}{r^2}}-\frac{2m}{r}, 
\end{equation}
\begin{equation}
    h(r)=\left(\sqrt{1-\frac{a^2}{r^2}}-\frac{2m}{r}\right)\left(1-\frac{a^2}{r^2}\right).\label{Poly_h}
\end{equation}
This solution has an event horizon at $r_h = \sqrt{a^2 + 4m^2}$ and, as in all previous cases, has a wormhole throat located at $r_t = a$, which is hidden behind the event horizon. It is interesting to note that there is no finite upper bound on $a$ that removes the event horizon in this model; therefore, no matter how large the radius of the wormhole throat becomes, the radius of the event horizon will always be large enough to conceal the throat.

As in the other cases, by choosing the coordinate transformation $r^2 = x^2 + a^2$, we can rewrite this spacetime as \eqref{metricageral_x} with
\begin{equation}
    f(x)=g(x)=\frac{x-2m}{\sqrt{x^2+a^2}}, \quad \Sigma(x)=\sqrt{x^2+a^2}.
\end{equation}
Although the spacetime is written in a very simple form in this coordinate system, for the purposes of this work, we will always consider the spacetimes expressed in the form \eqref{metricageral}.

Although there are many other BB models in the literature \cite{Lobo:2020ffi,Huang:2019arj}, in this article we will focus on the four models presented here, as they are among the most studied in various contexts.

\section{Geodesics}\label{SEC:geodesics}
Let us now turn our attention to the geodesic motion of neutral massive particles immersed in a spacetime described by a general line element written as in equation \eqref{metricageral}. Thus, we have that, for timelike geodesics
\begin{equation}
 \label{geodesicgeneral}   -1 = -f(r)\dot{t}^2 + \frac{\dot{r}^2}{h(r)} + r^2\dot{\theta}^2 + r^2\sin^2\theta\dot{\phi}^2,
\end{equation}
where the dot represents the derivative with respect to the proper time $\tau$. We will assume that the test bodies follow radial geodesics, so we will assume $\theta = \pi/2$, $\dot{\theta} = 0$, and $\dot{\phi} = 0$. Additionally, since we have a timelike Killing vector, we have the following conservation equation:
\begin{equation}
    E = f(r)\dot{t}.
\end{equation}

In the above equation, associated with the conservation of the energy of the particle along the geodesic motion, the constant $E$ represents the energy per unit mass of the test body. With this, we can rewrite equation \eqref{geodesicgeneral} as
\begin{eqnarray}
    -\frac{E^2}{f(r)} + \frac{\dot{r}^2}{h(r)} = -1,
\end{eqnarray}
or
\begin{eqnarray}
  \label{rgeodesica}  \frac{dr}{d\tau} = \pm \sqrt{\frac{[E^2 - f(r)]h(r)}{f(r)}}.
\end{eqnarray}

Since we will assume that the body is moving radially toward the BB, from now on we will only consider the negative sign of the equation above. Moreover, if the particle is released from rest at $r = b$, its energy per unit mass will then be given by $E = \sqrt{f(r = b)}$. With this, if the spacetime is asymptotically flat, then we have the special case where $E = \sqrt{f(b \to \infty)} = 1$.  

\section{Tidal forces}\label{SEC:TF}
In this section, we will present and discuss the main equations in the context of the study of tidal forces. First, our discussion will be based on the general line element given in \eqref{metricageral}, and then the derived equations will be applied to different specific cases.
\subsection{General equations}
As stated earlier, our main goal in this work is to investigate the tidal forces experienced by observers immersed in BB-like spacetimes. For this, we will make use of the geodesic deviation equation, which describes the relative acceleration between two infinitesimally close particles, given by
\begin{equation}
    \frac{D^2\xi^\mu}{D\tau^2} = K^\mu_{\;\;\;\gamma}\; \xi^\gamma,
\end{equation}
where $\xi^\mu$ is the infinitesimal displacement vector between nearby geodesics and $K^\mu_{\;\;\;\gamma}$ is the tidal tensor given in terms of Riemann tensor by
\begin{equation}
    K^\mu_{\;\;\;\gamma} = R^\mu_{\;\;\,\alpha\beta\gamma}u^{\alpha}u^\beta,
\end{equation}
where $u^\mu$ is the timelike vector tangent to the geodesic, which satisfies $u^\alpha u^\beta g_{\alpha\beta} = - 1$. To compute the components of the tidal force, we must use the tetrad formalism, that is, we need to employ the orthonormal basis set of tetrads to analyze the tidal force experienced by a test body in the vicinity of the BB. We assume that the test body follows a radial geodesic. Thus, the metric tensor written in terms of the tetrads is given by
\begin{equation}
    g^{\mu\nu} = \eta^{\hat{a}\hat{b}}\hat{e}_{\hat{a}}^{\;\; \mu}\hat{e}_{\hat{b}}^{\;\; \nu},
\end{equation}
where the hats indices are the tetrad basis indices, and $\eta_{\hat{a}\hat{b}} = \text{diag}(-1,\; 1,\; 1,\; 1)$ is the Minkowski metric in Cartesian coordinates.

To understand the effects of tidal forces, we need to observe how the spatial acceleration between two test bodies in free fall toward the BB evolves. In our case, we will consider that the observer is attached to the first particle, which follows a worldline given by $x^\mu(\tau)$ that passes through an event $P$ of the manifold at a given instant $\tau$. In this context, the tetrad basis will define the Instantaneous Rest Frame (IRF) of the first particle, that is, the observer, at $P$. The tetrad basis $\hat{e}_{\hat{a}}^{\;\;\mu}$ for a radially free-falling observer in the BB spacetime is then typically chosen so that the timelike vector aligns with the observer's 4-velocity. For an observer falling from rest, the tetrad can be written as:
\begin{eqnarray}
    \hat{e}_{\hat{0}}^{\;\; \mu}&=& \left(\frac{E}{f},\;-\sqrt{\frac{(E^2 - f)h}{f}},\;0,\;0\right),\\
    \hat{e}_{\hat{1}}^{\;\;\mu}&=& \left(-\frac{E^2 - f}{f},\;E\sqrt{\frac{h}{f}},\;0,\;0\right),\\
    \hat{e}_{\hat{2}}^{\;\;\mu}&=& \left(0,\;0,\;\frac{1}{r},\;0\right),\\
    \hat{e}_{\hat{3}}^{\;\;\mu}&=& \left(0,\;0,\;0,\;\frac{1}{r\sin\theta}\right).
\end{eqnarray}

\begin{figure}[htb]
    \centering
    \includegraphics[width=0.8\linewidth]{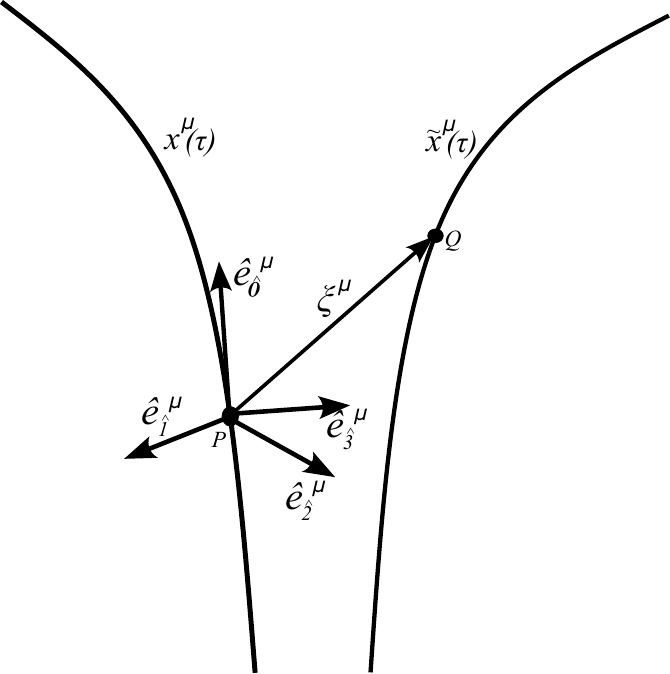}
    \caption{A schematic representation of the tetrad basis attached to the particle at $P$ is shown. The infinitesimal displacement vector $\xi^\mu$ between two nearby geodesics $x^\mu(\tau)$ and $\tilde{x}^\mu(\tau)$ is also depicted.}
    \label{esquema}
\end{figure}

In Fig. \ref{esquema}, we present a diagram representing the worldline of the first particle, to which the tetrad basis is attached. Furthermore, the figure also depicts the infinitesimal displacement vector between its nearby geodesics $x^\mu(\tau)$ and $\tilde{x}^\mu(\tau)$.

The timelike vector $\hat{e}_{\hat{0}}^{\;\; \mu}$ is the 4-velocity, i.e. $\hat{e}_{\hat{0}}^{\;\; \mu} = u^\mu$, and $ \hat{e}_{\hat{1}}^{\;\;\mu}, \hat{e}_{\hat{2}}^{\;\;\mu}$, and $\hat{e}_{\hat{3}}^{\;\;\mu}$ form an orthonormal basis for the three-dimensional space in the vicinity of the free-falling observer. $\hat{e}_{\hat{1}}^{\;\;\mu}$ points in the radial direction, describing variations in coordinate $r$, and is normalized and orthogonal to the temporal vector $\hat{e}_{\hat{0}}^{\;\; \mu}$. $\hat{e}_{\hat{2}}^{\;\; \mu}$ points in the polar $\theta$ direction, corresponding to small angular variations in latitude, while $\hat{e}_{\hat{3}}^{\;\; \mu}$ points in the azimuthal $\phi$ direction, describing angular variations around the symmetry axis. These vectors are mutually orthogonal and satisfy the Minkowski normalization condition 
\begin{equation}
    \hat{e}_{\hat{a}}^{\;\;\mu}\hat{e}_{\mu}^{\;\;\hat{b}} = \delta_{\hat{a}}^{\;\;\hat{b}},
\end{equation}
ensuring that a local observer measuring distances and angles in this frame obtains results consistent with special relativity. Computing the Riemann tensor associated with the metric \eqref{metricageral} and projecting it onto the tetrad basis of a free-falling observer in a BB spacetime, one can show that the tidal tensor in the tetrad frame is given by
\begin{equation}
    K^{\hat{a}}_{\;\;\hat{b}} = \text{diag}(0,k_1,k_2,k_2),
\end{equation}
where
\begin{eqnarray}
    k_1 &=& -\frac{h f''}{2 f}-\frac{f' h'}{4 f}+\frac{h f'^2}{4 f^2},\\
    k_2 &=& \frac{(E^2 - f)h'}{2rf}  - \frac{E^2hf'}{2rf^2}.
\end{eqnarray}
It is interesting to note that for solutions described by the line element \eqref{metricageral} with $f(r) = h(r)$, the tidal tensor component $k_2$ does not depend on the energy.

With the above equations, we can write the equations that describe the relative acceleration between two nearby observers as
\begin{eqnarray}
\label{tidal1}    \frac{D^2\xi^{\hat{r}}}{D\tau^2} &=& \left(-\frac{h f''}{2 f}-\frac{f' h'}{4 f}+\frac{h f'^2}{4 f^2}\right)\xi^{\hat{r}},\\
  \label{tidal2}  \frac{D^2\xi^{\hat{\theta}}}{D\tau^2} &=& \left(\frac{(E^2 - f)h'}{2rf}  - \frac{E^2hf'}{2rf^2}\right)\xi^{\hat{\theta}},\\
  \label{tidal3}   \frac{D^2\xi^{\hat{\phi}}}{D\tau^2} &=& \left(\frac{(E^2 - f)h'}{2rf}  - \frac{E^2hf'}{2rf^2}\right)\xi^{\hat{\phi}}.
\end{eqnarray}

The specific equations will depend on the type of BB spacetime considered, which will be analyzed case by case in the following subsection. However, before that, it is useful to write the equations that describe the variation of the displacement vector as a function of the coordinate $r$. This vector provides information about the deformation experienced by a test body in free fall within the BB spacetime. To obtain components \eqref{tidal1}-\eqref{tidal3}, we must recall the equation \eqref{rgeodesica}, which allows us to write the following second-order differential equations for the components of $\xi^{\hat{a}}$:
\begin{widetext}
\begin{eqnarray}
  &&  \left(\frac{(E^2 - f)h}{f}\right){\xi^{\hat{r}}}'' + \left(\frac{(E^2 - f)fh' - E^2hf'}{2f^2}\right){\xi^{\hat{r}}}' + \left(\frac{h f''}{2 f}+\frac{f' h'}{4 f}-\frac{h f'^2}{4 f^2}\right)\xi^{\hat{r}} = 0,\label{xi1}\\
     &&  \left(\frac{(E^2 - f)h}{f}\right){\xi^{\hat{\theta}}}'' + \left(\frac{(E^2 - f)fh' - E^2hf'}{2f^2}\right){\xi^{\hat{\theta}}}' - \left(\frac{(E^2 - f)h'}{2rf}  - \frac{E^2hf'}{2rf^2}\right)\xi^{\hat{\theta}} = 0,\label{xi2}\\
       &&  \left(\frac{(E^2 - f)h}{f}\right){\xi^{\hat{\phi}}}'' + \left(\frac{(E^2 - f)fh' - E^2hf'}{2f^2}\right){\xi^{\hat{\phi}}}' - \left(\frac{(E^2 - f)h'}{2rf}  - \frac{E^2hf'}{2rf^2}\right)\xi^{\hat{\phi}} = 0.\label{xi3}
\end{eqnarray}
\end{widetext}

The equations above generally cannot be solved analytically, so we must resort to numerical methods to solve them. Consequently, we need to establish initial conditions for their resolution. In our case, there are two initial conditions that are commonly used in the literature in the context of tidal force studies, namely:
\begin{eqnarray}
    \xi^{\hat{\beta}} (b) > 0, \;\; {\xi^{\hat{\beta}}} '(b) = 0, \;\;(ICI),\label{ICI}\\
       \xi^{\hat{\beta}}(b) = 0, \;\; {\xi^{\hat{\beta}}}'(b) > 0, \;\;(ICII),\label{{ICII}}
\end{eqnarray}
where $\xi^{\hat{\beta}}(b)$ is the $\hat{\beta} = \{\hat{r},\hat{\theta}, \hat{\phi}\}$ component of the infinitesimal displacement vector at $r = b$. Furthermore, condition (ICI) corresponds to the release from rest at $r = b$, a body composed of dust without internal interaction, while condition (ICII) corresponds to letting a body composed of dust explode at $r = b$. In this work, although we mention both conditions (ICI) and (ICII) for completeness, we restrict our analysis to (ICI), since it corresponds to the physically relevant situation of a body released from rest at $r=b$, which is the main focus of this study.

With the general important equations in hand, we can now perform a detailed analysis of them for different BB spacetimes.

\subsection{ Tidal Forces effects in SV spacetimes}
As a first application, we will now turn to the SV spacetime. With this, considering the equation \eqref{SVEq}, we can rewrite the equations \eqref{tidal1}, \eqref{tidal2}, and \eqref{tidal3} for this case as
\begin{eqnarray}
    \frac{D^2\xi^{\hat{r}}}{D\tau^2} &=& \frac{(2r^2 - 3a^2)m}{r^5}\; \xi^{\hat{r}},\label{k1_SV}\\
      \frac{D^2\xi^{\hat{\theta}}}{D\tau^2} &=& \frac{a^2[3m + r(E^2 - 1)] - mr^2}{r^5}\xi^{\hat{\theta}},\label{k2_SV}\\ 
      \frac{D^2\xi^{\hat{\phi}}}{D\tau^2} &=& \frac{a^2[3m + r(E^2 - 1)] - mr^2}{r^5}\xi^{\hat{\phi}}.\label{k3_SV}
\end{eqnarray}

We can draw some analytical conclusions from these equations. First, it is straightforward to see that the radial tidal force goes to zero at the point $r = r^r_0$
\begin{equation}
    r^r_0 = \sqrt{\frac{3}{2}}a,
\end{equation}
and reaches a maximum value at the point $r = r^r_{max}$
\begin{equation}
    r^r_{max} = \sqrt{\frac{5}{2}}a.
\end{equation}
If $a > 2\sqrt{2/3}m$, the point where the radial component of the tidal force is zero will be outside the event horizon, so in this case it will be possible, in principle, to observe this compression effect. For $a > 2\sqrt{2/5}m$, the maximum of the radial component will be located outside the event horizon. The angular tidal goes to zero at $r = r^a_0$
\begin{equation}
    r^a_0=\frac{a^2 \left(E^2-1\right)+\sqrt{a^4 \left(E^2-1\right)^2+12 a^2 m^2}}{2 m},
\end{equation}
and reaches a minimum value at the point $r^a_{min}$
\begin{equation}
    r^a_{min}=\frac{2 a^2 \left(E^2-1\right)+\sqrt{4 a^4 \left(E^2-1\right)^2+45 a^2 m^2}}{3 m}.
\end{equation}
We see that the expressions for the minimum point and for the point where the sign change occurs explicitly depend on the energy. These expressions can be significantly simplified if we consider the case $E=1$, where we obtain
\begin{equation}
    r^a_0=\sqrt{3} a, \quad \mbox{and} \quad r^a_{min}=\sqrt{5} a.
\end{equation}
For an arbitrary energy value, the point where the angular component reaches its minimum value or where it changes sign will be outside the horizon if $a$ assumes the values $a>2 m/\sqrt{2 E^2+1}$ and $a>2 \sqrt{3} m/\sqrt{8 E^2+7}$, respectively. These values can be simplified for the case where we choose $E=1$, and thus obtain $a>2m/\sqrt{3}$ and $a>2m/\sqrt{5}$.

We can also compute how these forces behave at the throat of the solution, $r=a$. In this case, we have
\begin{eqnarray}
     \frac{D^2\xi^{\hat{r}}}{D\tau^2}\Big|_{r = a} &=& -\frac{m}{a^3}\xi^{\hat{r}},\\
      \frac{D^2\xi^{\hat{\theta}}}{D\tau^2}\Big|_{r = a}  &=& \frac{2m + a(E^2 - 1)}{a^3}\xi^{\hat{\theta}},\\ 
      \frac{D^2\xi^{\hat{\phi}}}{D\tau^2}\Big|_{r = a}  &=& \frac{2m + a(E^2 - 1)}{a^3}\xi^{\hat{\phi}}.
\end{eqnarray}
All components are regular in the throat and become more intense as the value of $a$ decreases. In the Schwarzschild limit, $a \to 0$, these components diverge.

\begin{figure}[htbp]
    \centering
    \includegraphics[width=1\linewidth]{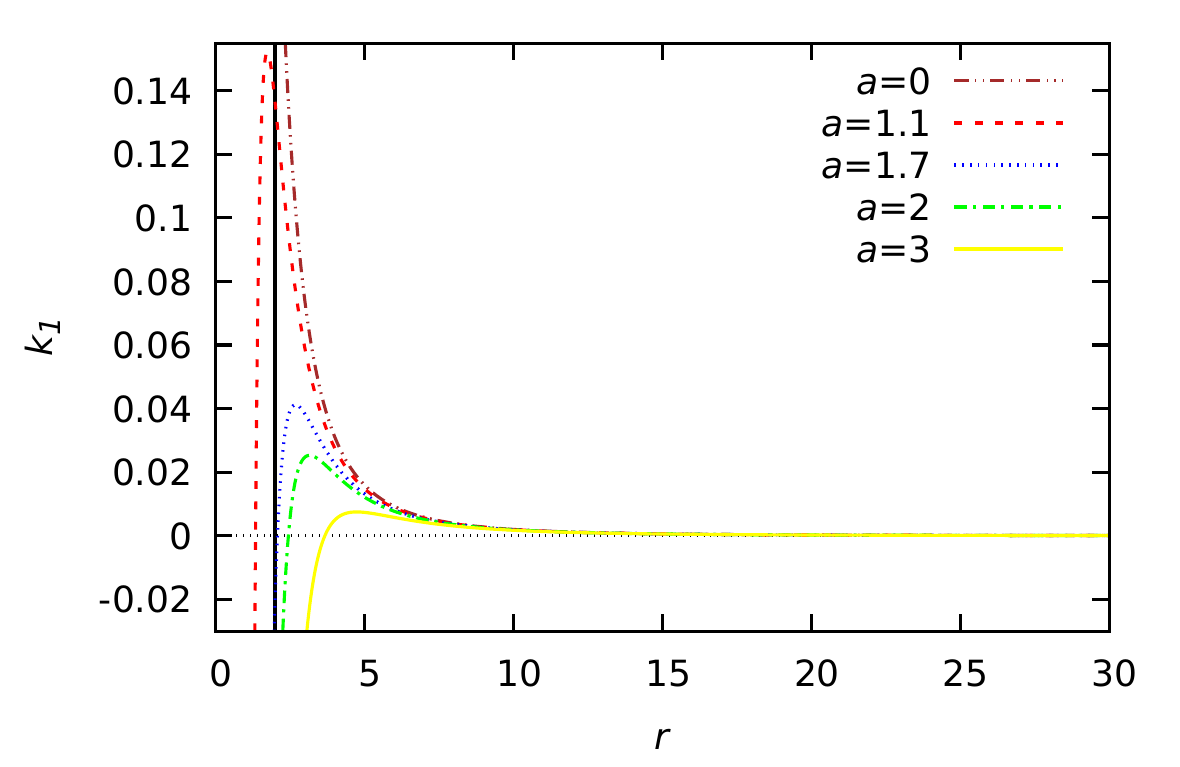}
    \caption{Radial tidal force as function of coordinate $r$ for different values of $a$, fixing $m = 1$. The vertical line is located at $r = 2m$.}
    \label{fig:k1_SV}
\end{figure}

\begin{figure}[htbp]
    \centering
    \includegraphics[width=1\linewidth]{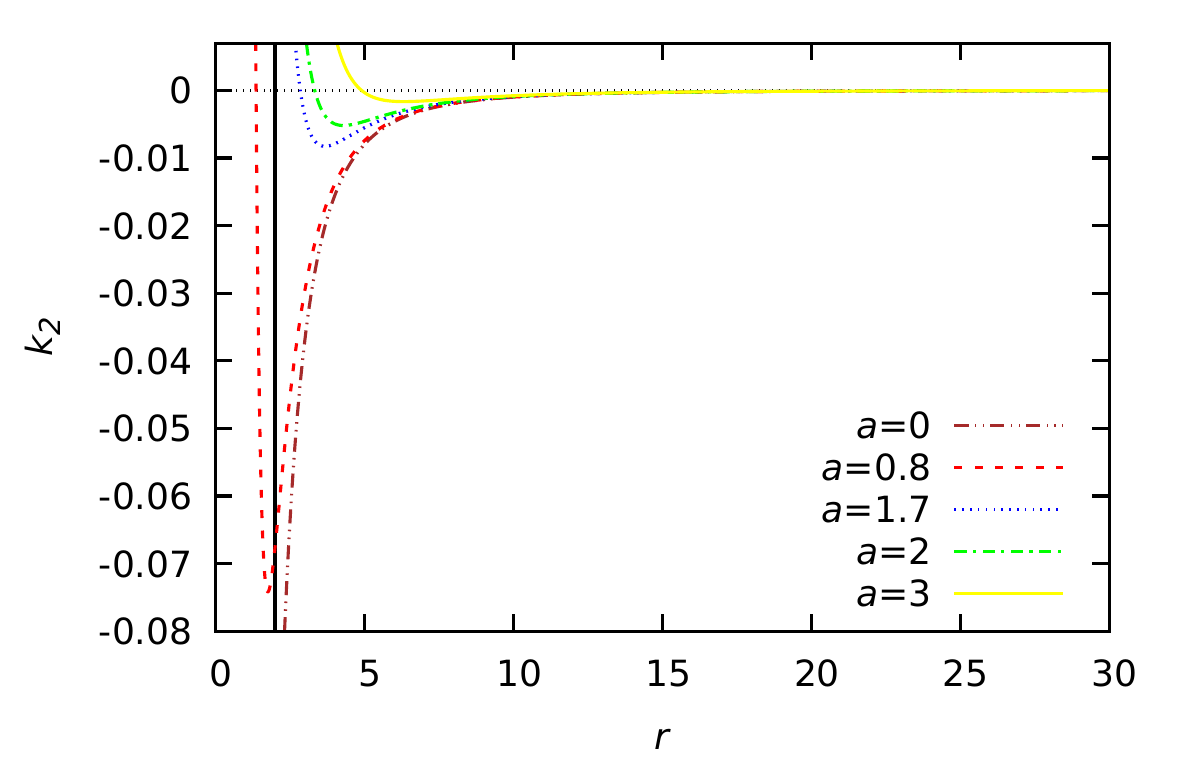}
    \caption{Angular tidal force as function of coordinate $r$ for different values of $a$, fixing $m = 1$ and $b = 30m$. The vertical line is located at $r = 2m$.}
    \label{fig:k2_SV}
\end{figure}

In Fig. \ref{fig:k1_SV}, we observe the behavior of the radial component of the tidal force, $k_1$. As discussed, we can see that if the value of the parameter $a$ is small enough, both the point where $k_1$ reaches its maximum and the point where $k_1$ changes sign remain hidden behind the event horizon. In Fig. \ref{fig:k2_SV}, we observe the behavior of the angular component of the tidal force, $k_2$. Similarly to what happens with the radial component, if the value of $a$ is small enough, the minimum points of $k_2$ and the point where $k_2$ changes sign are hidden by the event horizon, making their observation impossible. In general, for more distant points, $k_1$ takes positive values while $k_2$ takes negative values, leading to a stretching of the bodies in these regions. For smaller distances, $k_1$ takes negative values while $k_2$ takes positive values, resulting in a compression event.

Regarding the displacement vector for the SV case, we can attempt to write analytical expressions for the solutions of equations \eqref{xi1}-\eqref{xi3}. However, these expressions are quite extensive and written in terms of elliptic integrals. We can try to simplify by choosing the special case where $E \to 1$. In this case, we find that
\begin{equation}
   \xi^{\hat{r}}=\frac{a_1}{\sqrt{r}}+ \frac{2}{5} a_2 r \left(\sqrt{r^2-a^2}+a\,i  \, _2F_1\left(\frac{1}{2},\frac{3}{4};\frac{7}{4};\frac{r^2}{a^2}\right)\right),
\end{equation}
\begin{equation}
   \xi^{\hat{\theta}}= b_1 r+\frac{2 i b_2 r^{3/2} \, _2F_1\left(\frac{1}{4},\frac{1}{2};\frac{5}{4};\frac{r^2}{a^2}\right)}{a},
\end{equation}
\begin{equation}
   \xi^{\hat{\phi}}= b_1 r+\frac{2 i b_2 r^{3/2} \, _2F_1\left(\frac{1}{4},\frac{1}{2};\frac{5}{4};\frac{r^2}{a^2}\right)}{a}.
\end{equation}
 where $\, _2F_1\left(x_1,x_2;x_3;x_4\right)$ is the hypergeometric function and the integration constants $a_1$, $a_2$, $b_1$, and $b_2$ can be determined by imposing the boundary conditions at $b$. These analytical expressions represent the special case where $b \to \infty$. Next, we shall analyze graphically the behavior of the displacement vector by considering a fixed value of $b = 100m$.

\begin{figure}[htbp]
    \centering
    \includegraphics[width=1\linewidth]{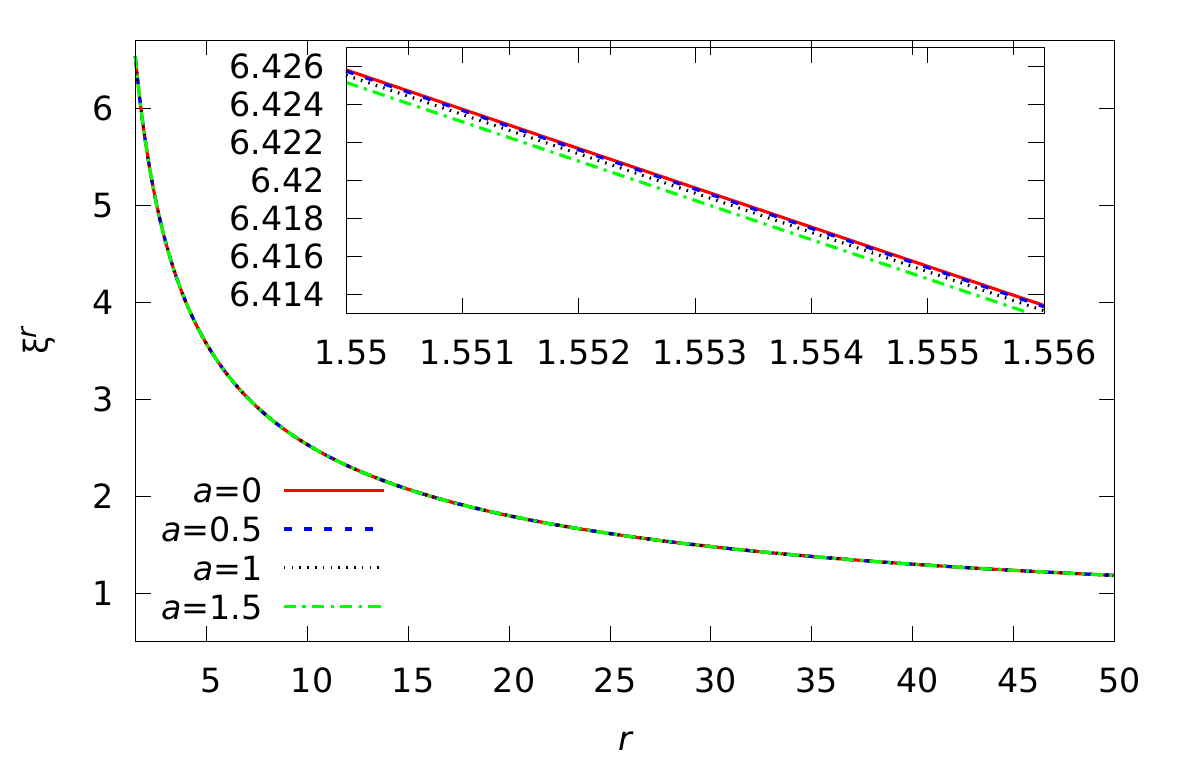}
    \includegraphics[width=1\linewidth]{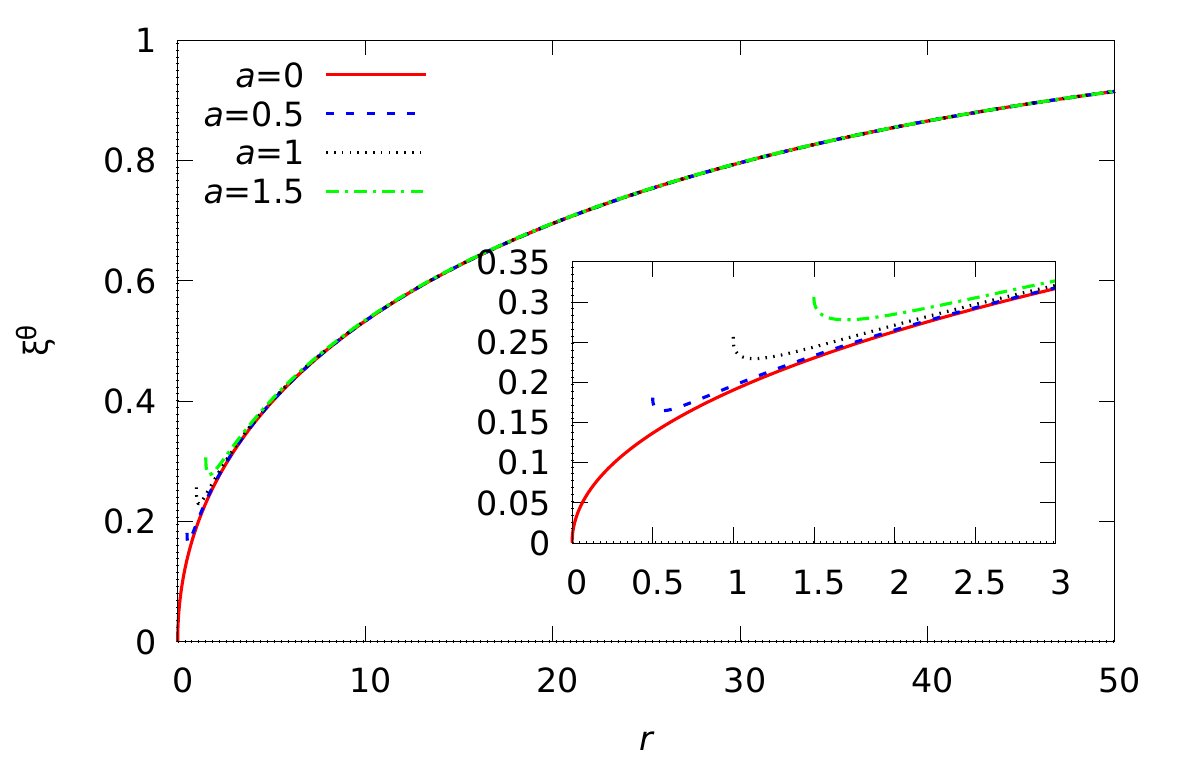}
    \caption{Radial (top) and tangential (bottom) components of the displacement vector for radial geodesics in the SV spacetime as a function of the radial coordinate for different values of $a$. In this case, the boundary conditions \eqref{ICI} were used with $b=100$ and $m=1$.}
    \label{fig:n_SV}
\end{figure}

\begin{figure}[htbp]
    \centering
    \includegraphics[width=1\linewidth]{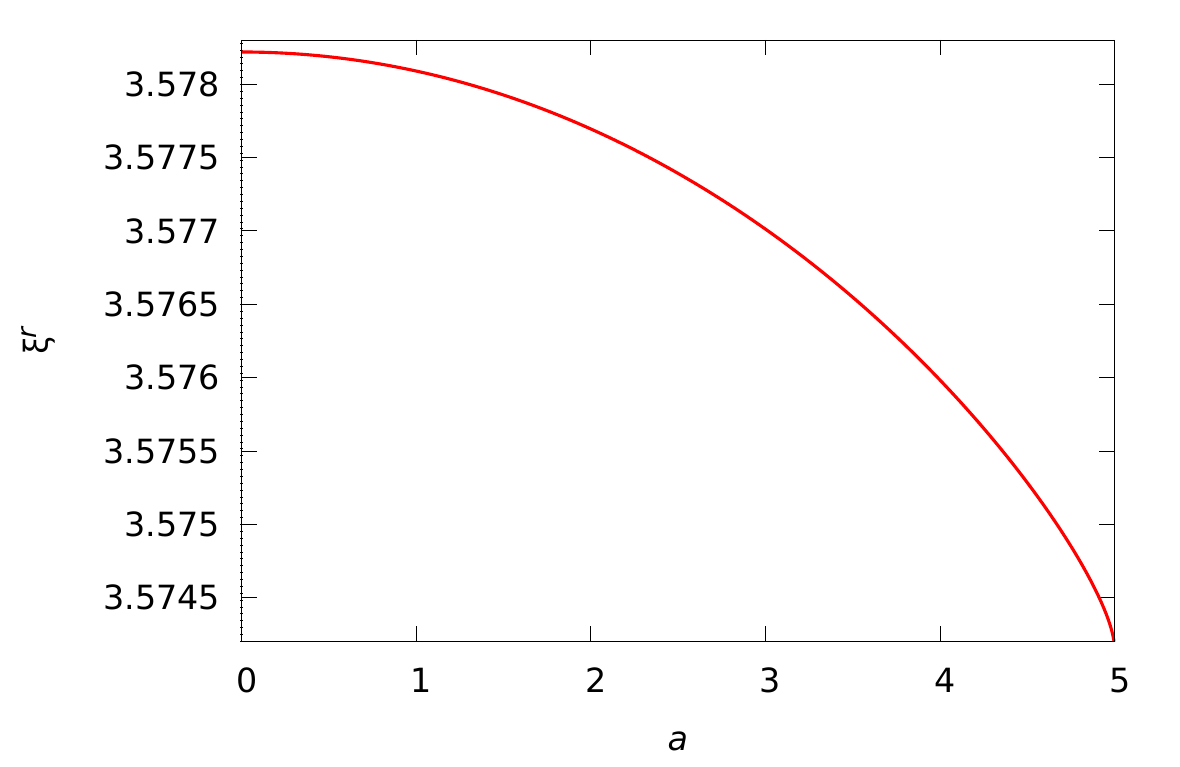}
    \includegraphics[width=1\linewidth]{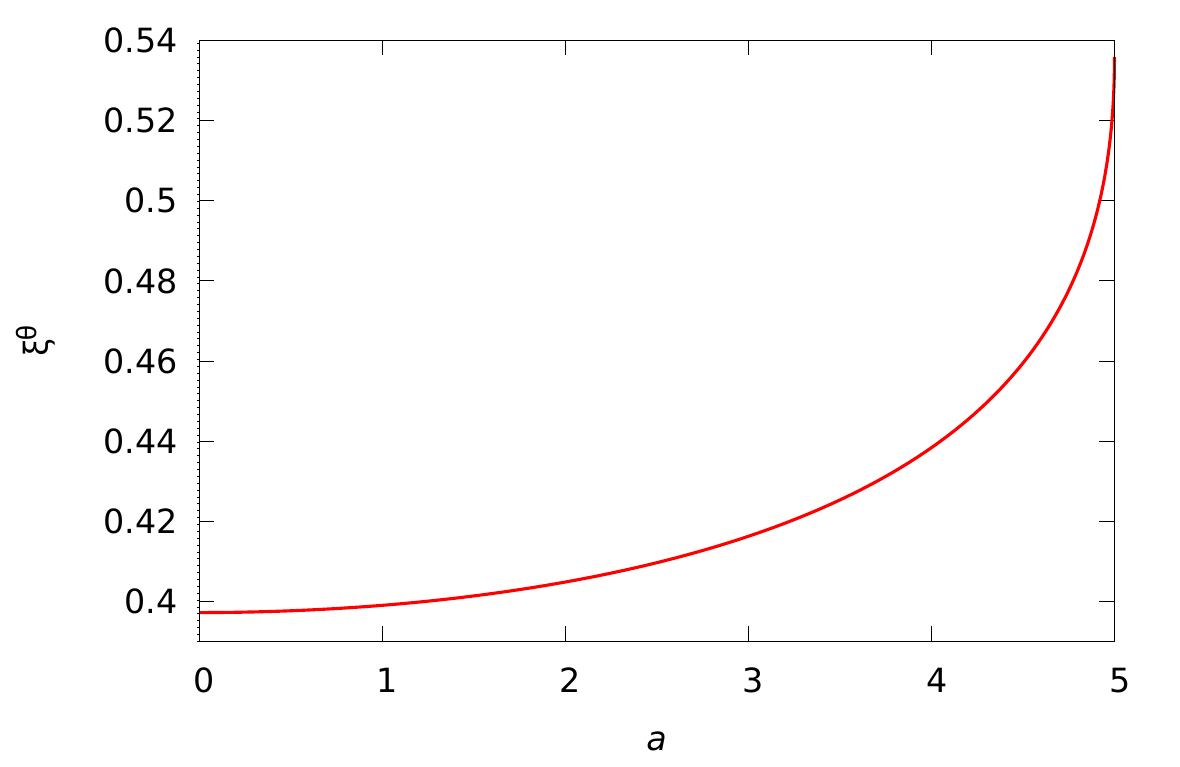}
    \caption{Radial (top) and tangential (bottom) components of the displacement vector for radial geodesics in the SV spacetime as a function of the parameter $a$ for a specific value of the radial coordinate $r=5$. In this case, the boundary conditions \eqref{ICI} were used with $b=100$ and $m=1$.}
    \label{fig:n_a_SV}
\end{figure}

In Figure \ref{fig:n_SV}, we observe the behavior of the radial and angular components of the displacement vector for radial geodesics in the SV spacetime considering the initial condition \eqref{ICI}. Regarding the radial component, we notice that changing the regularization parameter $a$ does not significantly alter the shape of the curve. Furthermore, for a given point, the radial component is larger for smaller values of $a$. This result is expected, as the radial acceleration is greater for smaller values of $a$, leading to a smaller separation for larger values of the regularization parameter. For the angular component, the behavior differs. Although the differences are small, the angular separation increases with increasing values of $a$. Furthermore, it is important to note that, for nonzero $a$, some curves abruptly terminate. This occurs because, in this coordinate system, the allowed region is $a<r<+\infty$, which means that the plots end at the wormhole throat. In Figure \ref{fig:n_a_SV}, we fix the point at which we analyze the components of the displacement vector and examine how they behave for different values of $a$. We observe that the differences in the radial component are quite small, whereas in the angular component, they become more noticeable only for higher values of $a$.

Although they may appear different, our results are consistent with those presented in the reference \cite{Arora:2023ltv}. The apparent discrepancy arises due to the choice of coordinate system, and this section demonstrates how different coordinate systems can affect the visualization of the results.

\subsection{Tidal Forces effects in Bardeen-type BB spacetimes}
Now we will analyze the effects of the tidal force in the background given by a Bardeen-type BB. 

Taking into account the equations \eqref{bardeef} and \eqref{bardeenh}, we can rewrite the equations \eqref{tidal1}, \eqref{tidal2}, and \eqref{tidal3} for this case as
\begin{eqnarray}
    \frac{D^2\xi^{\hat{r}}}{D\tau^2} &=& \frac{m \left(15 a^4-15 a^2 r^2+2 r^4\right)}{r^7}\xi^{\hat{r}},\label{k1_Bardeen}\\
      \frac{D^2\xi^{\hat{\theta}}}{D\tau^2} &=& \frac{-5 a^4 m+a^2 r^2 \left(6 m+\left(E^2-1\right) r\right)-m r^4}{r^7}\xi^{\hat{\theta}},\nonumber\\ \label{k2_Bardeen}\\ 
      \frac{D^2\xi^{\hat{\phi}}}{D\tau^2} &=& \frac{-5 a^4 m+a^2 r^2 \left(6 m+\left(E^2-1\right) r\right)-m r^4}{r^7}\xi^{\hat{\phi}}.\nonumber\\ \label{k3_Bardeen}
\end{eqnarray}

Regarding the radial force, it is possible to conclude that it goes to zero at two different points of the radial coordinate, given by
\begin{equation}
  \label{zero1r}  {r_0^r}^{(1)}= \frac{a}{2}\sqrt{15 + \sqrt{105}},\quad
   {r_0^r}^{(2)} = \frac{a}{2}\sqrt{15 - \sqrt{105}},
\end{equation}
such that we can conclude that, unlike the SV case, here the radial tidal force changes sign twice. Depending on the value of $a$, the first sign change that $k_1$ undergoes may be located outside the event horizon. However, for cases where horizons exist, the second sign change will always be hidden behind the horizons.

The radial force reaches its maximum value at $r = a$, as will be better understood later. Furthermore, it has a local maximum and a global minimum, both located at
\begin{equation}
 \label{localmaxr}   {r_{max}^r}^{local} = \frac{a}{2}\sqrt{25 + \sqrt{345}},\quad
     r_{min}^r = \frac{a}{2}\sqrt{25 - \sqrt{345}}.
\end{equation}
For cases where horizons exist, the minimum point will always be hidden behind the horizon, while the local maximum point will only be hidden behind the event horizon for small values of $a$.

With respect to the angular force, it also goes to zero at two different values of the radial coordinate. However, for arbitrary values of $E$, these expressions become extremely complicated. Thus, for the specific case where $E \to 1$, the points where the angular tidal force goes to zero are given by
\begin{equation}
    {r_0^a}^{(1)} = \sqrt{5}a,\quad   {r_0^a}^{(2)} = a.
\end{equation}
It is interesting to note that for $E \to 1$, the angular tidal force goes to zero at $r \to a$, while the radial tidal force reaches its maximum at the same point. This behavior differs significantly from the SV case. Similarly as in the case of the radial component, depending on the values of the parameter $a$, the first sign change may be hidden behind the horizon or occur outside it, whereas the second sign change is always hidden behind the horizon in the cases where the event horizon exists.

The angular tidal force has a local minimum and a global maximum, both given by
\begin{equation}
    {r_{min}^a}^{local} = a \sqrt{5 + 2\sqrt{\frac{10}{3}}},\quad
     r_{max}^a = a \sqrt{5 - 2\sqrt{\frac{10}{3}}}.
\end{equation}
In cases where horizons exist, the maximum point will always be hidden behind the event horizon, while the minimum point may be located outside the event horizon for certain values of $a$.

The components of the tidal forces calculated at the minimum of the radial coordinate are given by
\begin{eqnarray}
     \frac{D^2\xi^{\hat{r}}}{D\tau^2}\Big|_{r = a} &=& \frac{2m}{a^3}\xi^{\hat{r}},\\
      \frac{D^2\xi^{\hat{\theta}}}{D\tau^2}\Big|_{r = a}  &=& \frac{E^2 - 1}{a^3}\xi^{\hat{\theta}},\\ 
      \frac{D^2\xi^{\hat{\phi}}}{D\tau^2}\Big|_{r = a}  &=& \frac{E^2 - 1}{a^3}\xi^{\hat{\phi}}.
\end{eqnarray}

As mentioned above, it can be observed that for $E \to 1$, the angular tidal force goes to zero at the location of the throat.

\begin{figure}[htbp]
    \centering
    \includegraphics[width=1\linewidth]{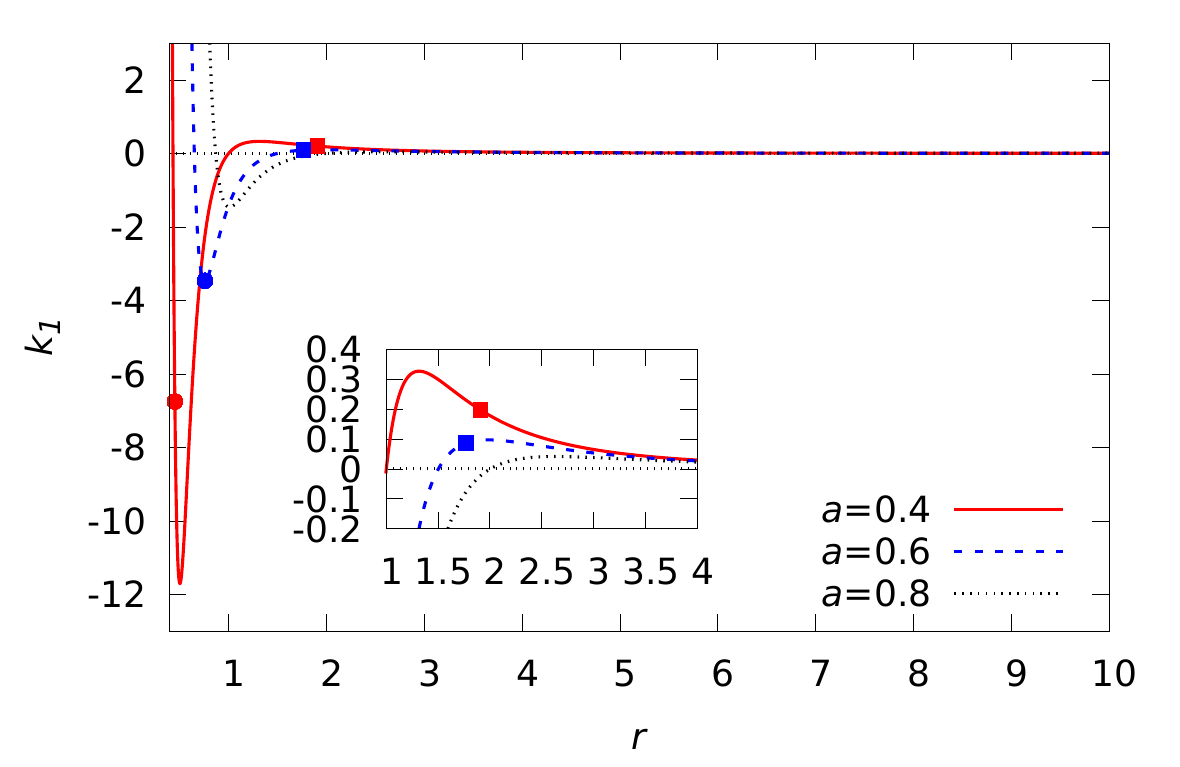}
    \caption{Radial tidal force for the Bardeen-type BB as function of coordinate $r$ for different values of $a$, fixing $m = 1$. The square points on the curves represent the position of the event horizon, while the circular points represent the position of the Cauchy horizon.}
    \label{fig:k1_B}
\end{figure}

In Fig. \ref{fig:k1_B}, we show the behavior of the radial tidal force as a function of $r$. The plot reveals that the radial force changes sign twice. For sufficiently large values of $r$, it takes positive values, reaching a local maximum at the position given by \eqref{localmaxr}, after which it begins to decrease until it reaches zero at the position given by \eqref{zero1r}. Beyond this point, it becomes negative, reaching its minimum value at \eqref{localmaxr}, where it starts increasing again until it reaches zero once more at \eqref{zero1r}. From there, it turns positive again to the throat, located at $r=a$. Furthermore, for the chosen parameter values, these sign changes in the radial tidal force are initially hidden by the event horizon.

\begin{figure}[htbp]
    \centering
    \includegraphics[width=1\linewidth]{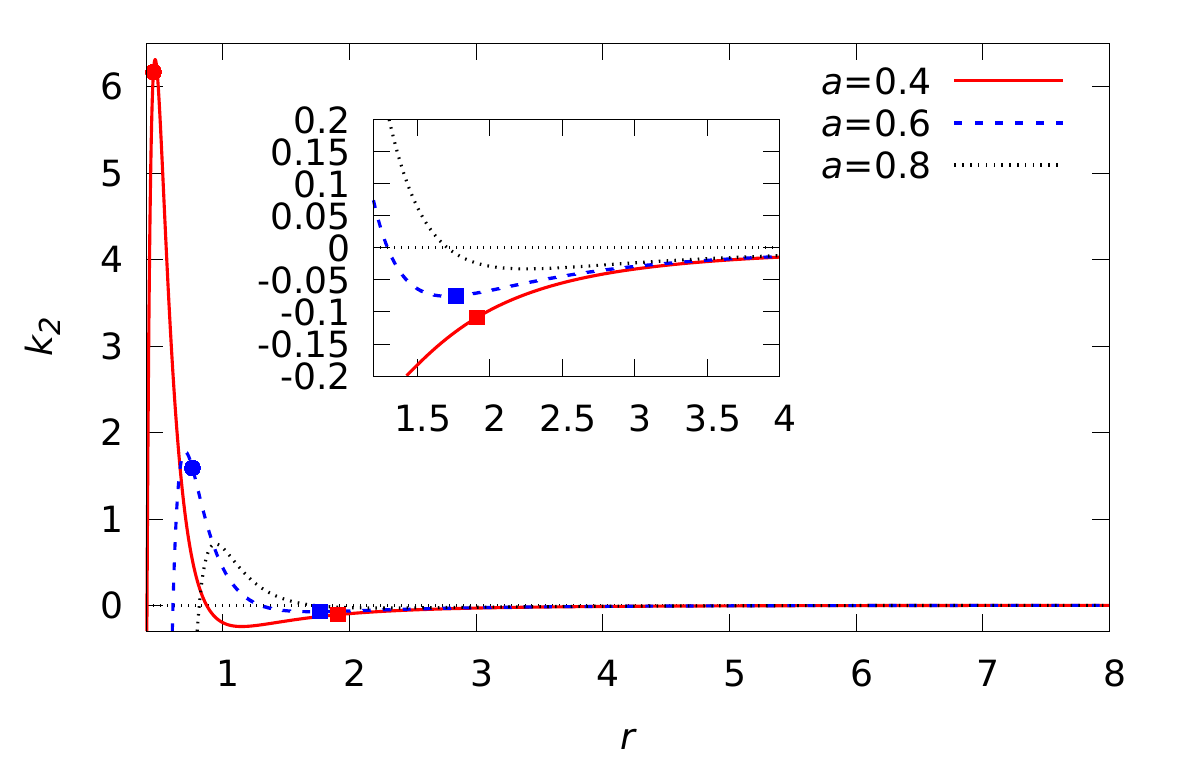}
    \caption{Angular tidal force for the Bardeen-type BB as function of coordinate $r$ for different values of $a$, fixing $m = 1$ and $b = 10m$. The square points on the curves represent the position of the event horizon, while the circular points represent the position of the Cauchy horizon.}
    \label{fig:k2_B}
\end{figure}

Regarding the displacement vector for the Bardeen-type case, it is not possible to solve \eqref{xi1}-\eqref{xi3} analytically for arbitrary values of the energy. However, we can try to simplify by choosing the special case where $E \to 1$. In this case, we find that
\begin{eqnarray}
   \xi^{\hat{r}}&=&\frac{c_1 \sqrt{r^2-a^2}}{r^{3/2}}+c_2 \left(\frac{
   36 a^2 r^2+4 r^4-45 a^4}{10 r\sqrt{r^2-a^2 }}\right.\nonumber\\
    &-&\left.\frac{9 a^{5/2}\sqrt{r^2-a^2 }}{4 r^{3/2}}\left( \tan ^{-1}\sqrt{\frac{r}{a}}+ \tanh ^{-1}\sqrt{\frac{r}{a}}\right)\right),\nonumber\\
\end{eqnarray}
\begin{equation}
   \xi^{\hat{\theta}}=d_1 r+\frac{d_2 r}{\sqrt{a}} \left(\tan ^{-1}\sqrt{\frac{r}{a}}-\tanh ^{-1}\sqrt{\frac{r}{a}}\right),
\end{equation}
\begin{equation}
   \xi^{\hat{\phi}}=d_1 r+\frac{d_2 r}{\sqrt{a}} \left(\tan ^{-1}\sqrt{\frac{r}{a}}-\tanh ^{-1}\sqrt{\frac{r}{a}}\right),
\end{equation}
where the integration constants $c_1$, $c_2$, $d_1$, and $d_2$ can be determined by imposing the boundary conditions in $b$. Although the expressions are not written in terms of special functions, as in the SV case, the analytical expressions are neither clear nor informative. Once again, these analytical expressions represent the special case where $b \to \infty$ and next, we shall analyze graphically the behavior of the displacement vector by considering a fixed value of $b = 200m$.

\begin{figure}[htbp]
    \centering
    \includegraphics[width=1\linewidth]{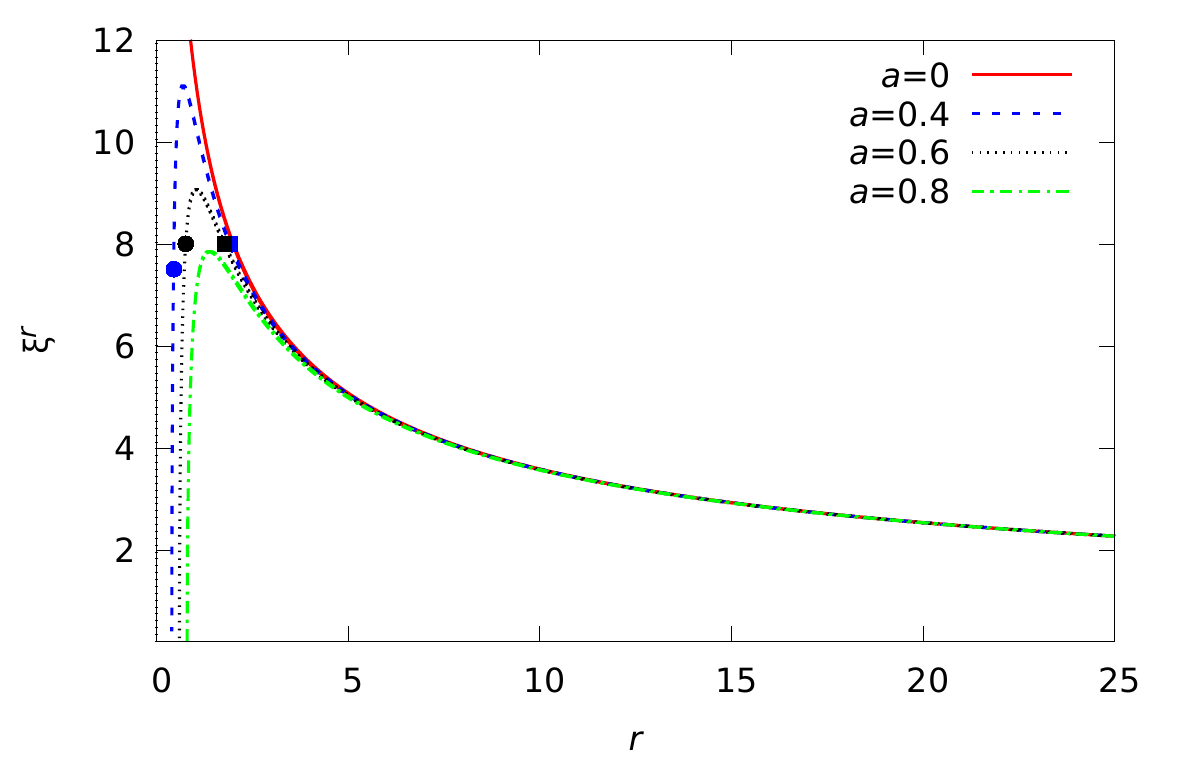}
    \includegraphics[width=1\linewidth]{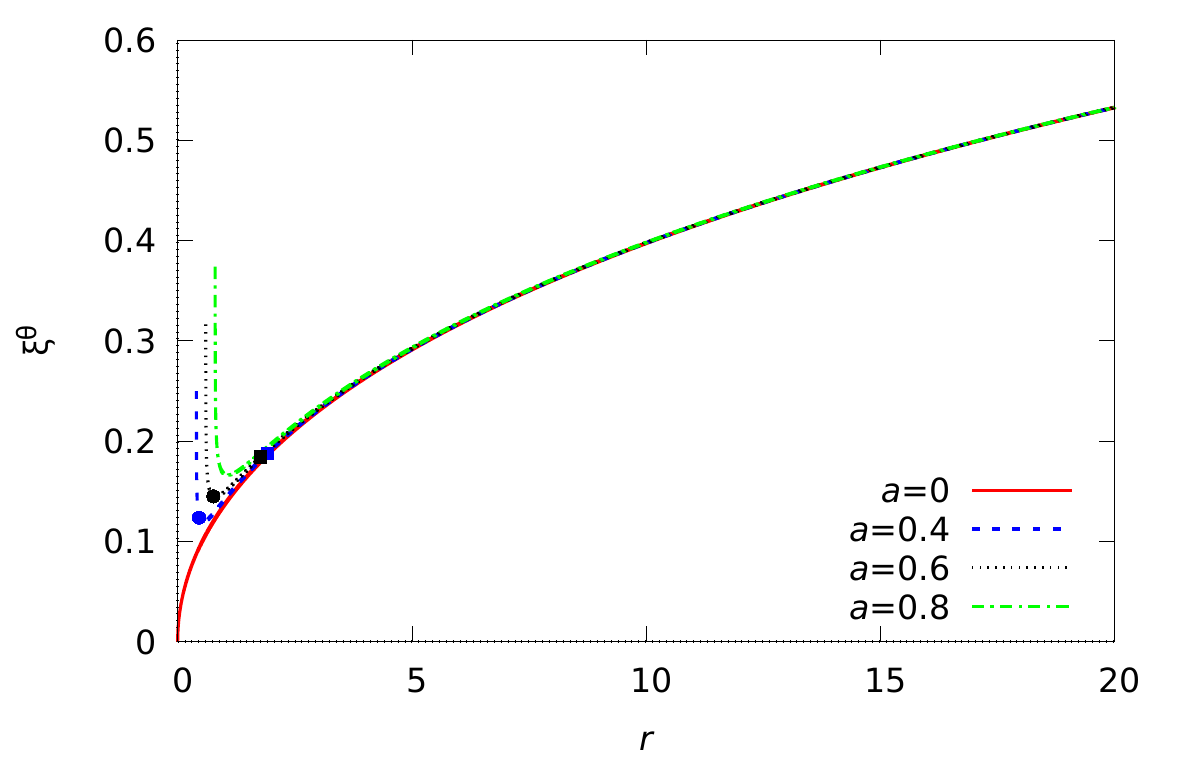}
    \caption{Radial (top) and tangential (bottom) components of the displacement vector for radial geodesics in the Bardeen-type spacetime as a function of the radial coordinate for different values of $a$. In this case, the boundary conditions \eqref{ICI} were used with $b=200$ and $m=1$. The square points on the curves represent the position of the event horizon, while the circular points represent the position of the Cauchy horizon. The positions of the horizons are shown only for the BB cases.}
    \label{fig:n_BT}
\end{figure}

\begin{figure*}[htbp]
    \centering
    \includegraphics[width=.9\linewidth]{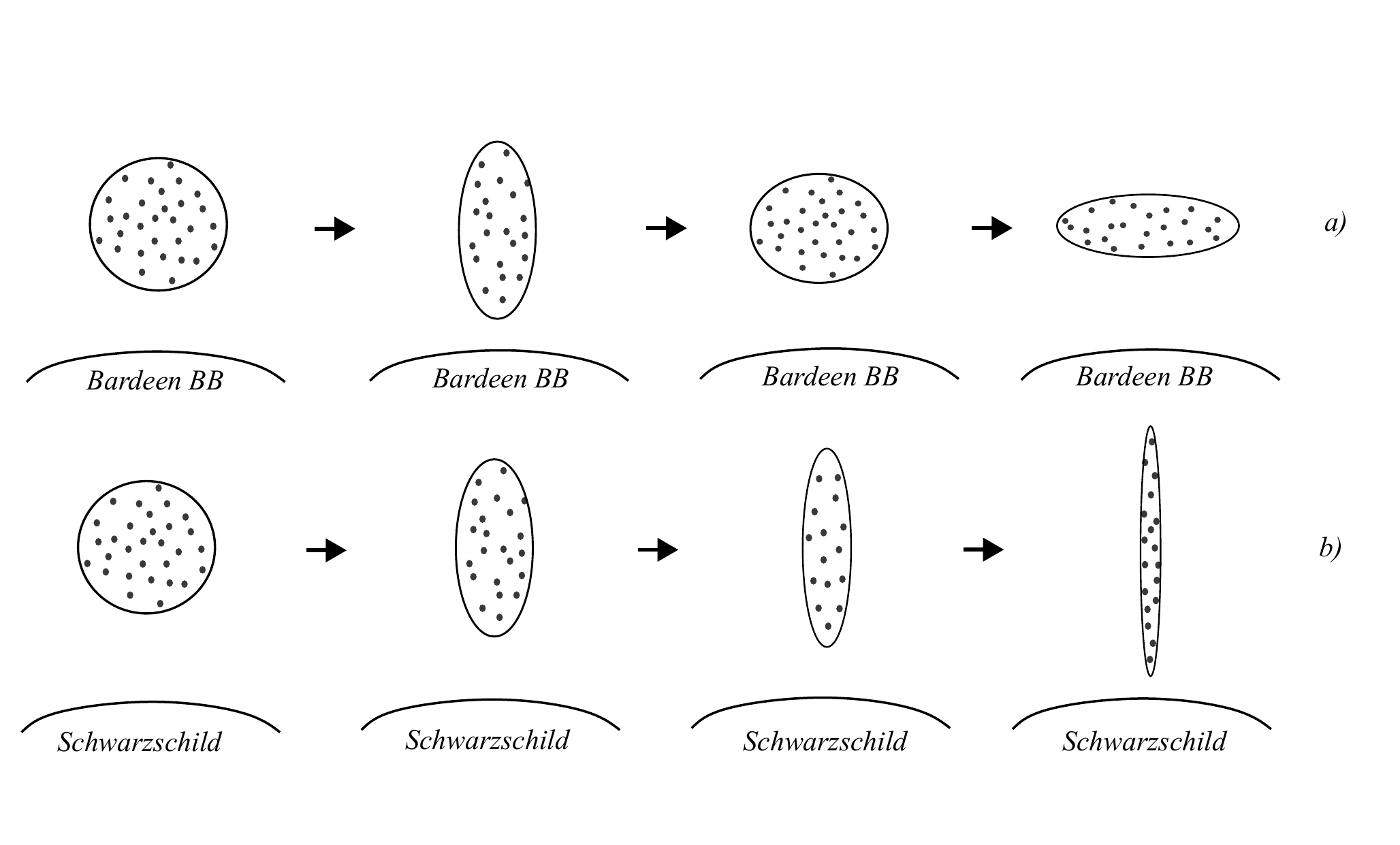}
    \caption{Schematic representation of the stretching and compression phenomena that a body experiences in the vicinity of a Bardeen-type BB (a), compared to the Schwarzschild case (b), where only stretching occurs.}
    \label{fig:Queda_SV}
\end{figure*}

In Fig. \ref{fig:n_BT}, we show the behavior of the radial and angular components of the displacement vector. The differences between the Bardeen-type spacetime and the Schwarzschild spacetime become quite evident, in contrast to what we observed in the SV case. It is clear that there are regions where the radial component is compressed rather than stretched and the angular component is stretched rather than compressed. As a result, in more internal regions, we find a compression effect on the bodies. This effect does not occur in the Schwarzschild case. However, it is hidden by the presence of an event horizon in the cases where horizons exist. This phenomenon is illustrated in Fig. \ref{fig:Queda_SV}, where we observe that there are regions in which the bodies undergo stretching and regions in which they undergo compression.

\subsection{Tidal Forces effects in holonomy corrected BH}
We now proceed to study the behavior of tidal forces for the holonomy corrected model.

Considering equation \eqref{HCEq}, we can rewrite equations \eqref{tidal1}, \eqref{tidal2}, and \eqref{tidal3} for this case as
\begin{eqnarray}
    \frac{D^2\xi^{\hat{r}}}{D\tau^2} &=& \frac{(4r - 5a)m}{2r^4}\; \xi^{\hat{r}},\label{k1_LQG}\\
      \frac{D^2\xi^{\hat{\theta}}}{D\tau^2} &=& \frac{(4a - 2r)m + ar(E^2 - 1)}{2r^4}\xi^{\hat{\theta}},\label{k2_LQG}\\ 
      \frac{D^2\xi^{\hat{\phi}}}{D\tau^2} &=& \frac{(4a - 2r)m + ar(E^2 - 1)}{2r^4} \xi^{\hat{\phi}}.\label{k3_LQG}
\end{eqnarray}
The radial tidal force for this case reaches zero at the location given by $r_0^r = 5a/4$, while its maximum value is reached at $r_{max}^r = 5a/3$. For $a < 6m/5$, the maximum point of the radial component is hidden by the event horizon, while the point where $k_1 = 0$ is hidden by the horizon for $a < 8m/5$.

With respect to the angular part of the tidal force, it reaches zero and a minimum value at
\begin{eqnarray}
    r_0^a &=& \frac{4ma}{2m -a(E^2 - 1)},\\
    r_{min}^a &=& \frac{16ma}{3(2m -a(E^2 - 1))}.
\end{eqnarray}
For $a < 2m/(1 + E^2)$, the point where $k_2$ vanishes is hidden by the event horizon, while for $a < 6m/(5 + 3E^2)$, the point where $k_2$ reaches its minimum value is also hidden by the event horizon.

We can also analyze the behavior of the tidal forces at the wormhole throat, which are given by:
\begin{equation}
     \frac{D^2\xi^{\hat{r}}}{D\tau^2}\Big|_{r = a} =-\frac{m}{2a^3}\xi^{\hat{r}},
\end{equation}
\begin{equation}
     \frac{D^2\xi^{\hat{\theta}}}{D\tau^2}\Big|_{r = a}=\frac{2m + a(E^2 - 1)}{2a^3}\xi^{\hat{\theta}}, 
\end{equation}
\begin{equation}
    \frac{D^2\xi^{\hat{\phi}}}{D\tau^2}\Big|_{r = a}=\frac{2m + a(E^2 - 1)}{2a^3}\xi^{\hat{\phi}}.
\end{equation}
These results, at least at the throat, are quite similar to the SV case. More precisely, the magnitude of these components is half the value of the components in the SV case.

\begin{figure}[htbp]
    \centering
    \includegraphics[width=1\linewidth]{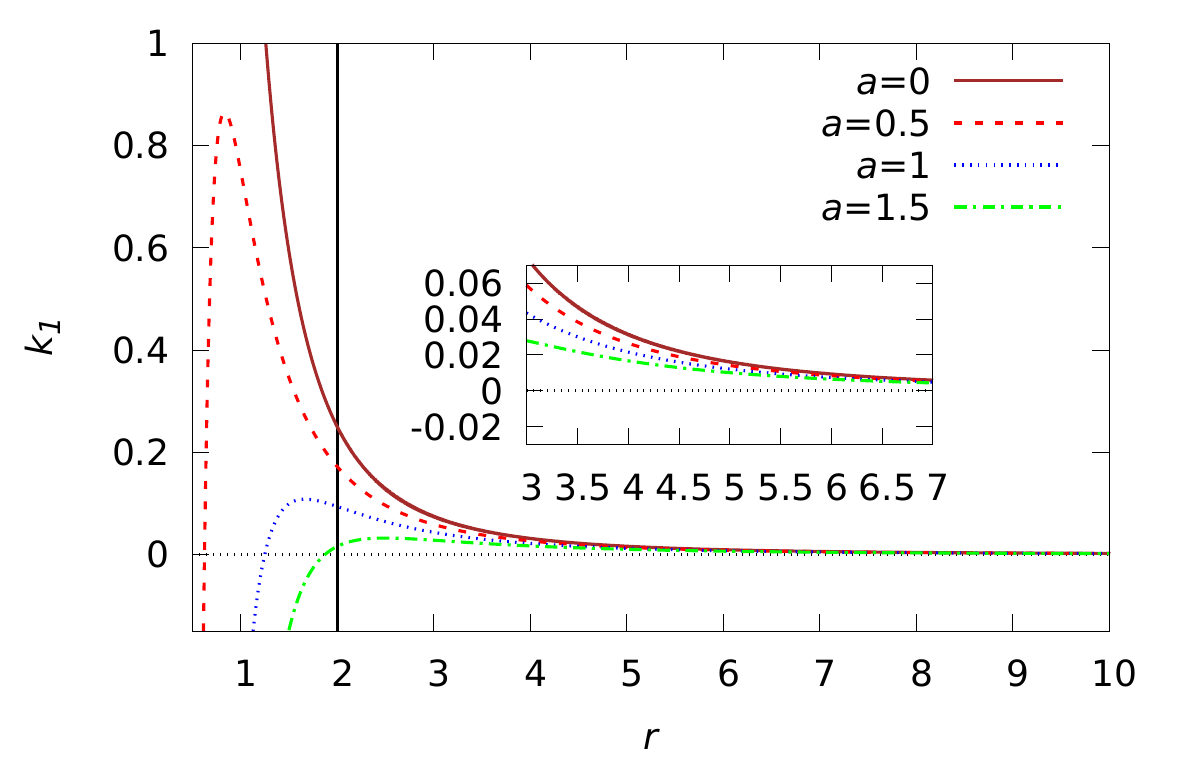}
    \caption{Radial tidal force for the holonomy corrected Schwarzschild BH as function of coordinate $r$ for different values of $a$, fixing $m = 1$. The vertical line is located at $r = 2m$.}
    \label{fig:k1_LQG}
\end{figure}

In Fig. \ref{fig:k1_LQG}, we show the behavior of the radial component of the tidal force, $k_1$, as a function of the radial coordinate. We observe that for distant points, $k_1$ is positive for all chosen values of $a$. In the Schwarzschild case, $k_1$ never changes sign and diverges at the origin, whereas for cases with $a \neq 0$, there is always a maximum positive value that $k_1$ can reach, after which it changes sign. This sign change may lead to a compression effect instead of a stretching effect. These points will only be exposed, that is, located outside the event horizon, for sufficiently large values of $a$, close to the limiting value this parameter can take.

\begin{figure}[htbp]
    \centering
    \includegraphics[width=1\linewidth]{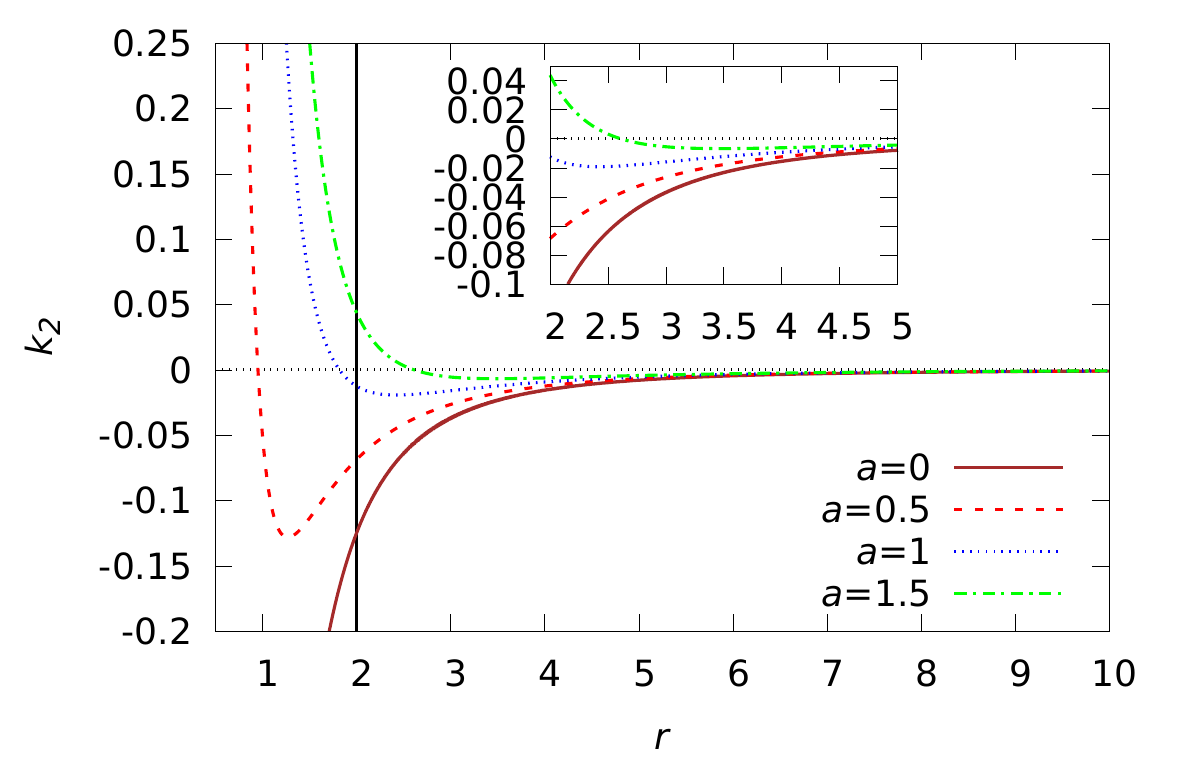}
    \caption{Angular tidal force for the holonomy corrected Schwarzschild BH as function of coordinate $r$ for different values of $a$, fixing $m = 1$ and $b=10$. The vertical line is located at $r = 2m$.}
    \label{fig:k2_LQG}
\end{figure}

In Fig. \ref{fig:k2_LQG}, we show the behavior of the angular component of the tidal force, $k_2$, as a function of the radial coordinate. The effects are similar to those observed in previous cases, where for $a \neq 0$, $k_2$ always reaches a negative minimum value and then changes sign, becoming positive. In the case $a = 0$, which corresponds to the Schwarzschild spacetime, this sign change does not occur. Once again, these points, which differ from the Schwarzschild case, may be hidden behind the event horizon depending on the value of the parameter $a$. Together with the result observed for $k_1$, these effects can lead to a compression phenomenon instead of a stretching effect.

Despite the sign change in $k_1$ and $k_2$, there is no guarantee that a transition from stretching to compression actually takes place. What may happen instead is simply a weaker stretching effect compared to the Schwarzschild case. To confirm whether the compression effect truly occurs, it is necessary to analyze the behavior of the displacement vector.

For arbitrary values of energy $E$, equations \eqref{xi1}--\eqref{xi3} can be solved analytically within the model \eqref{HCEq}, without the need for special functions. However, the expression for $\xi^r$ is rather complex. Therefore, we will consider the specific case $E=1$, which yields:
    \begin{equation}
   \xi^{\hat{r}}=\frac{a_1}{\sqrt{r}}+\frac{2}{15} a_2 \sqrt{1-\frac{a}{r}} \left(8 a^2+4 a r+3 r^2\right),
\end{equation}
\begin{equation}
   \xi^{\hat{\theta}}=b_1 r+\frac{2 b_2 r \tan ^{-1}\left(\sqrt{\frac{r-a}{a}}\right)}{\sqrt{a}},
\end{equation}
\begin{equation}
   \xi^{\hat{\phi}}=b_1 r+\frac{2 b_2 r \tan ^{-1}\left(\sqrt{\frac{r-a}{a}}\right)}{\sqrt{a}},
\end{equation}
where the constants $a_1$, $a_2$, $b_1$, and $b_2$ are determined by imposing the appropriate boundary conditions. Next, we plot the behavior of the displacement vector considering the value $b = 200m$.

\begin{figure}[htbp]
    \centering
    \includegraphics[width=1\linewidth]{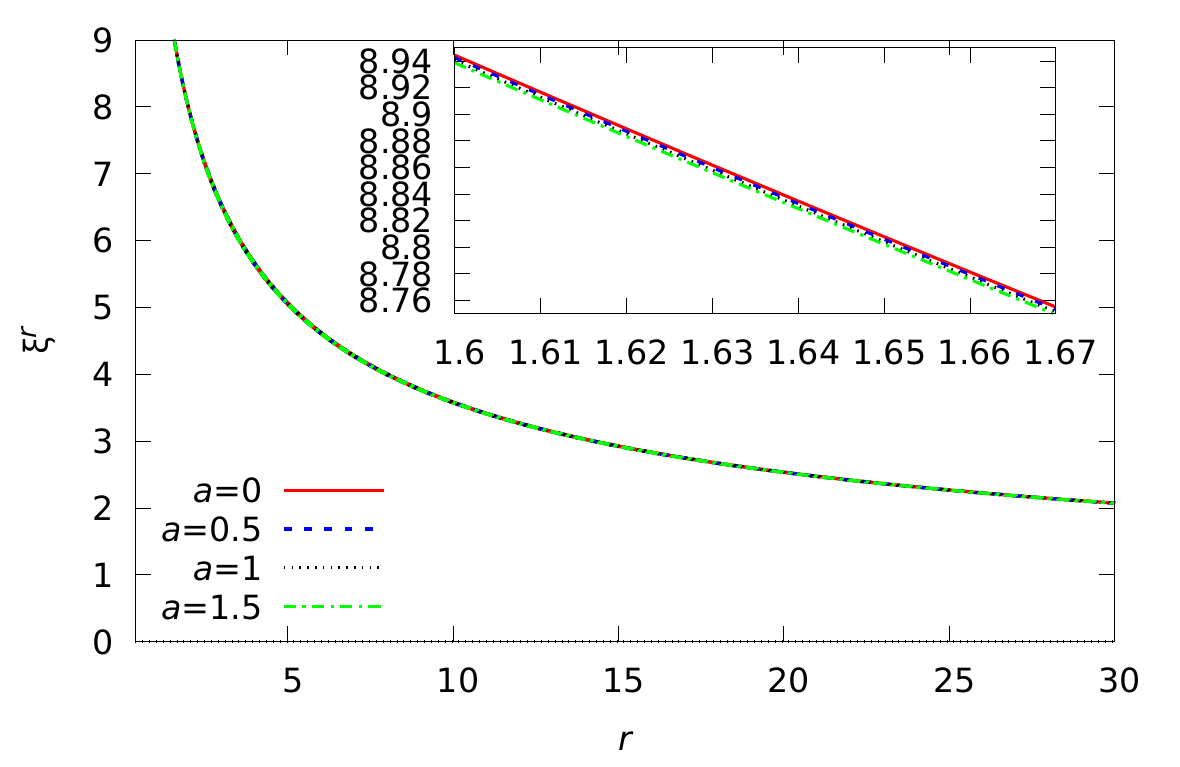}
    \includegraphics[width=1\linewidth]{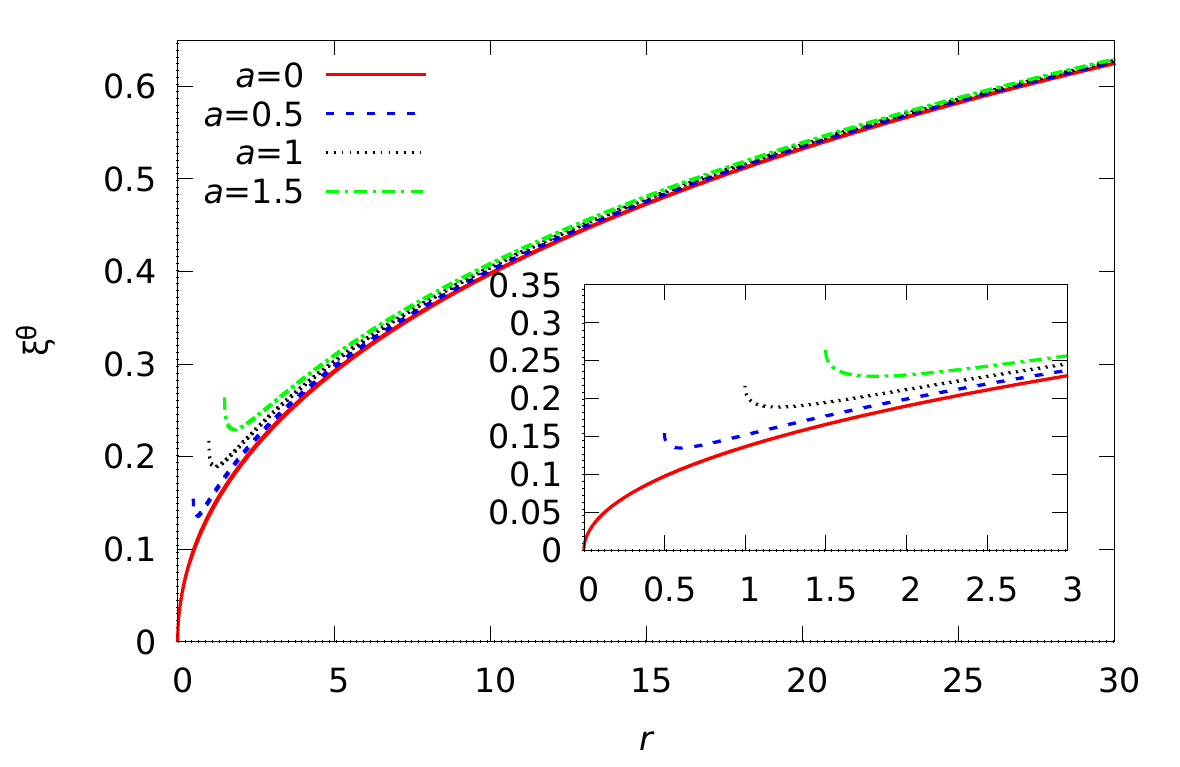}
    \caption{Radial (top) and tangential (bottom) components of the displacement vector for radial geodesics in the Holonomy corrected Schwarzschild BH as a function of the radial coordinate for different values of $a$. In this case, the boundary conditions \eqref{ICI} were used with $b=200$ and $m=1$.}
    \label{fig:n_LQG}
\end{figure}

In Fig. \ref{fig:n_LQG}, we analyze the behavior of the displacement vector for different values of the parameter $a$. We observe that for the holonomy corrected Schwarzschild BH, the displacement vector behaves very similarly to the SV case. The variation of $a$ does not significantly affect the radial component, $\xi^r$, while the largest changes in the angular component occur near the throat of the wormhole, which remains hidden behind the event horizon.

\subsection{Tidal Forces effects in Polymerized BH}
In the final case, we will compute the tidal forces for the polymerized BH. Considering equations \eqref{Poly_f} and \eqref{Poly_h} in equations \eqref{tidal1}, \eqref{tidal2}, and \eqref{tidal3}, we find
\begin{eqnarray}
    \frac{D^2\xi^{\hat{r}}}{D\tau^2} &=&\frac{4mr^2 + 3a^2(\sqrt{r^2 - a^2} - 2m)}{2r^5}\xi^{\hat{r}},\\
     \frac{D^2\xi^{\hat{\theta}}}{D\tau^2} &=&\frac{a^2(6m -3\sqrt{r^2 -a^2} + 2rE^2) -2mr^2}{2r^5}\xi^{\hat{\theta}},\\
     \frac{D^2\xi^{\hat{\phi}}}{D\tau^2} &=&\frac{a^2(6m -3\sqrt{r^2 -a^2} + 2rE^2) -2mr^2}{2r^5}\xi^{\hat{\phi}}.
\end{eqnarray}
The radial tidal force reaches zero at
\begin{equation}
    r_0^r=\frac{a}{4 m}\sqrt{\frac{3}{2}} \sqrt{3 a^2+16 m^2-a \sqrt{9 a^2+32 m^2}},
\end{equation}
This point will always remain hidden behind the event horizon. We can also compute the maximum point, but the analytical expression is quite lengthy. Depending on the chosen value of the parameter $a$, the maximum point can be located inside or outside the event horizon. For the angular component, the analytical expressions for the extremal points or the points where $k_2=0$ are even more complicated. In general, these points may be hidden behind the event horizon or not, depending on the value of $a$, just as in the case of $k_1$.

We can also analyze the behavior of the tidal forces at the wormhole throat, which are given by:
\begin{equation}
     \frac{D^2\xi^{\hat{r}}}{D\tau^2}\Big|_{r = a} =-\frac{m}{a^3}\xi^{\hat{r}},
\end{equation}
\begin{equation}
     \frac{D^2\xi^{\hat{\theta}}}{D\tau^2}\Big|_{r = a}=\frac{a E^2+2 m}{a^3}\xi^{\hat{\theta}}, 
\end{equation}
\begin{equation}
    \frac{D^2\xi^{\hat{\phi}}}{D\tau^2}\Big|_{r = a}=\frac{a E^2+2 m}{a^3}\xi^{\hat{\phi}}.
\end{equation}
At the wormhole throat, the radial component has the same expression as in the SV case, whereas the difference in the angular component resembles a modification in the particle's energy term.

\begin{figure}[htbp]
    \centering
    \includegraphics[width=1\linewidth]{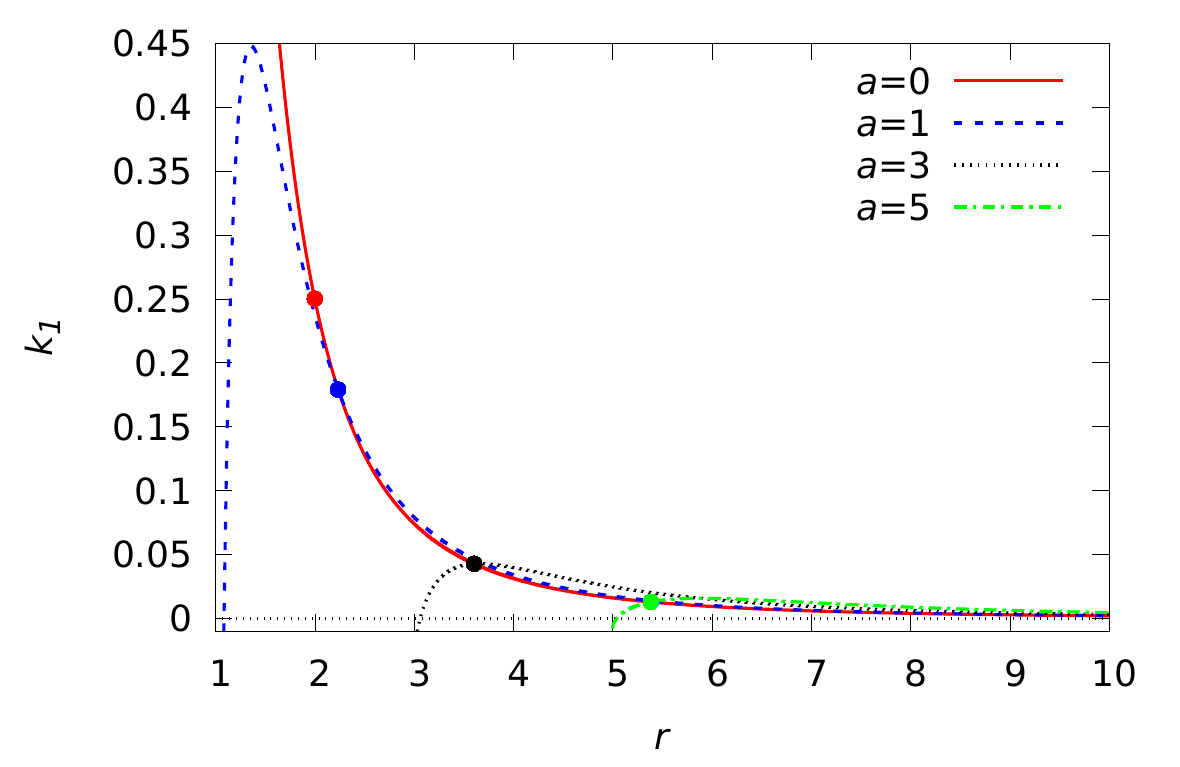}
    \caption{Radial tidal force for the polymerized BH as function of coordinate $r$ for different values of $a$, fixing $m = 1$. The circular points represent the position of the event horizon.}
    \label{fig:k1_poly}
\end{figure}

\begin{figure}[htbp]
    \centering
    \includegraphics[width=1\linewidth]{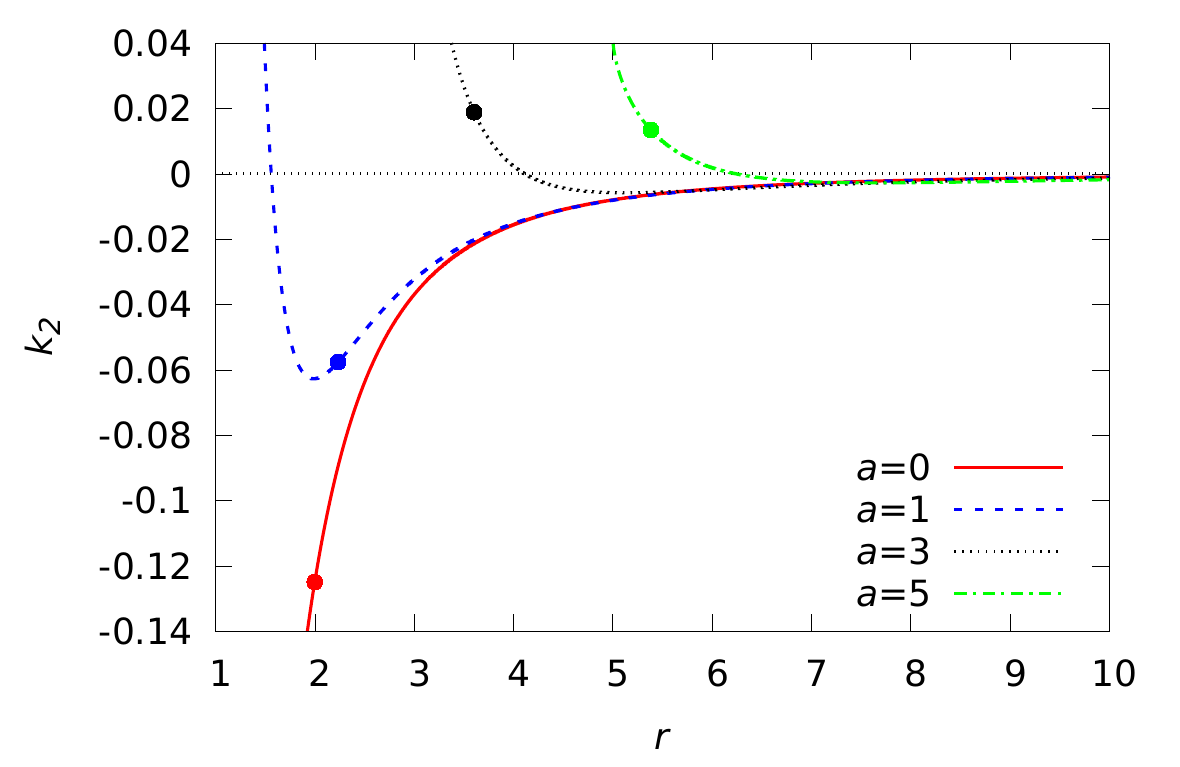}
    \caption{Angular tidal force for the polymerized BH as function of coordinate $r$ for different values of $a$, fixing $m = 1$ and $b = 10m$. The circular points represent the position of the event horizon.}
    \label{fig:k2_poly}
\end{figure}

In Figs. \ref{fig:k1_poly} and \ref{fig:k2_poly}, we show the behavior of the radial and tangential components, respectively, as functions of the radial coordinate for different values of $a$. Although no further analytical insights can be drawn, we observe that the sign change in $k_2$ may occur outside the event horizon, while the sign change in $k_1$ remains hidden. For both components, the extremum point may be concealed depending on the value of $a$.

\begin{figure}[htbp]
    \centering
    \includegraphics[width=1\linewidth]{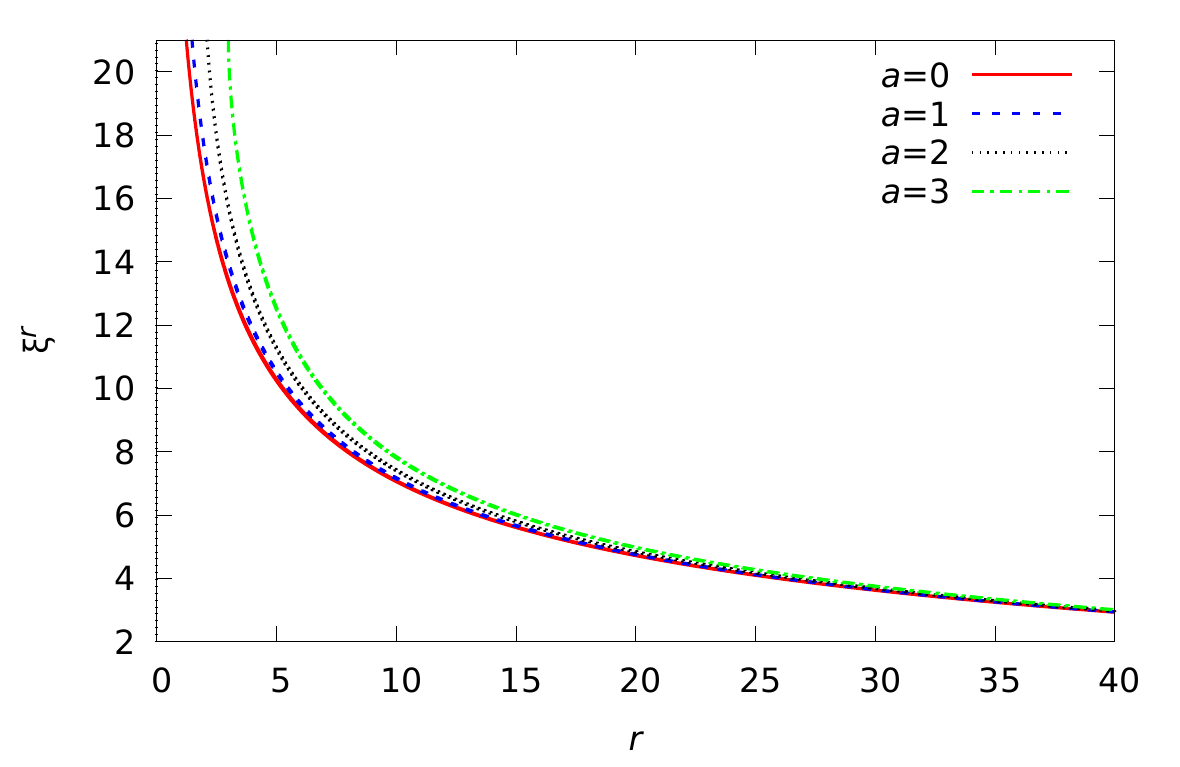}
    \includegraphics[width=1\linewidth]{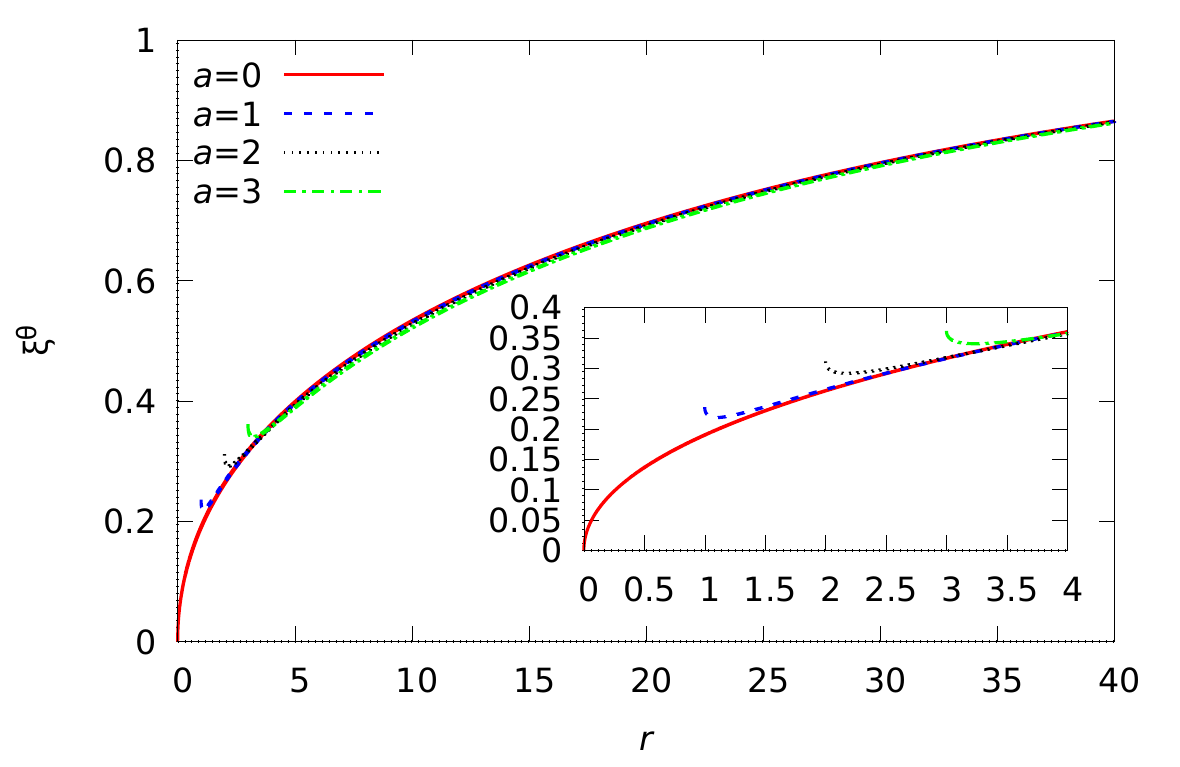}
    \caption{Radial (top) and tangential (bottom) components of the displacement vector for radial geodesics in the polymerized BH as a function of the radial coordinate for different values of $a$. In this case, the boundary conditions \eqref{ICI} were used with $b=100$ and $m=1$.}
    \label{fig:n_poly}
\end{figure}

With respect to the displacement vector, the situation becomes more complex for this spacetime. Due to the square-root term in the metric components, it is not possible to obtain analytical solutions, even in the limit $E \to 1$. We numerically analyzed the behavior of these components in Fig. \ref{fig:n_poly}. It is interesting to note that, at a given point, the radial component of the displacement vector becomes larger as the value of $a$ increases. This behavior is not observed in the cases previously discussed. The tangential component shows a similar trend, decreasing as $a$ increases, except in regions very close to the wormhole throat. In the previous cases, the radial component of the displacement vector always decreased as $a$ increased, while the tangential component increased with $a$, even in regions far from the wormhole throat.

\section{Conclusions}\label{SEC:conclusions}

In this work, our main objective was to investigate the effects of the tidal forces experienced by a test body in free fall on different backgrounds of BB. To this end, we used the so-called geodesic deviation equation, which describes the separation between two infinitesimally close particles. Using the tetrad formalism, we compute the components of the tidal tensor in the IRF for the general line element given in \eqref{metricageral}.

After expressing the main equations for the tidal forces and the infinitesimal displacement vector in terms of the metric functions $f(r)$ and $h(r)$, we turned our attention to the four BB spacetimes considered in this work, namely: SV, Bardeen-type BB, the holonomy corrected Schwarzschild spacetime, and the polymerized BH. For each case, we analyze the behavior of the tidal tensor components in the IRF, identifying possible divergences and regions of regularity. The comparison between the different models allowed us to highlight the physical features of each geometry as well as their impact on the tidal effects near the center of the configuration.
One of the main results is that, unlike the Schwarzschild spacetime, no divergences are observed in the three cases considered here. 

Regarding the SV case, an important result of our analysis is that the location of characteristic points of the tidal force components, such as where they vanish, reach extrema, or change sign, can lie outside the event horizon, depending on the value of the parameter $a$. As illustrated in Fig. \ref{fig:k1_SV}, small values of $a$ cause both the maximum of $k_1$ and the point where it changes sign to lie within the event horizon, hiding these features from an external observer. The same behavior is seen for the angular component $k_2$, as shown in Fig. \ref{fig:k2_SV}, with its minimum and sign change point also hidden when $a$ is small. 

The analysis of the components of the displacement vector allowed us to explore how the regularization parameter $a$ influences the separation between two nearby geodesics in the SV spacetime. We found that variations in $a$ have little impact on the radial component, which decreases slightly as $a$ increases, consistent with the fact that the radial acceleration weakens with larger $a$. On the other hand, the angular component shows a mild but noticeable growth for increasing $a$, especially at higher values. Despite small differences in appearance compared to the results in the literature, particularly \cite{Arora:2023ltv}, our findings remain consistent when considering differences in coordinate systems.

In contrast to the SV case, the Bardeen-like BB spacetime presents a richer tidal structure. While in the SV geometry the radial tidal force changes sign only once, here it changes sign twice. As shown in Fig. \ref{fig:k1_B}, depending on the value of the regularization parameter $a$, the first sign change can occur outside the event horizon, making it, in principle, observable. However, the second sign change and the local minimum of the radial force always remain hidden behind the horizon in geometries where a horizon exists. This indicates that, even though both spacetimes regularize the central singularity, the tidal environment near the core can be considerably more complex in the Bardeen-type case. Interestingly, for $E \to 1$, the radial tidal force reaches its maximum precisely in the throat $r=a$, a behavior that does not occur in the SV case.

The angular tidal force also exhibits different features. While in the SV case it vanishes only once, here it goes to zero at two different radii, and it has both a local minimum and a global maximum, as illustrated in Fig. \ref{fig:k2_B}. For $E \to 1$, the force vanishes exactly at the throat, where the radial force reaches its maximum, further highlighting the structural differences between the two geometries. Moreover, the location of the extrema and the changes in sign of the angular component are generally more sensitive to the value of $a$, and depending on its value, some of these characteristics may lie inside or outside the horizon.

The analysis of the displacement vector in Bardeen-type geometry highlights important qualitative differences when compared to the Schwarzschild and SV cases. Although the SV spacetime exhibited only minor deviations from the Schwarzschild behavior, the Bardeen-type solution reveals a much more intricate structure, with distinct regions of tidal compression and stretching. In particular, the radial component of the displacement vector shows compression in inner regions, while the angular component simultaneously exhibits stretching, an effect absent in the standard Schwarzschild case. Although these features are generally hidden behind the event horizon when it exists, their presence points to fundamental differences in the tidal environment.

The third case considered was the holonomy corrected Schwarzschild spacetime. This case turned out to be qualitatively very similar to the SV case. In this scenario, for large values of the parameter $a$, close to the maximum limit $a$, it is always possible for the sign changes in the radial and angular components of the tidal forces to occur outside the event horizon, as shown in the corresponding Figs. \ref{fig:k1_LQG} and \ref{fig:k2_LQG}. We also find that, in the holonomy corrected Schwarzschild case, the displacement vector exhibits qualitatively similar behavior to the SV spacetime. Variations in the parameter $a$ have little impact on the radial component, while changes in the angular component are more noticeable near the throat, still hidden behind the event horizon.

In the final case, we analyzed the polymerized BH. This case exhibits results that are quite different from the other models. The point where $k_1$ changes sign is always hidden behind the event horizon, while for $k_2$, this point may lie outside the horizon. It can be verified that, in certain regions, $k_1$ increases with $a$, whereas in the previous models, $k_1$ decreased with $a$. For this model, it was not possible to find analytical expressions for the displacement vector, even in the case $E=1$. It is also observed that, in a given region, the displacement vector increases with $a$, which again differs from the behavior found in the previous cases.

Therefore, our results show that the tidal forces experienced by a test body in a BB background differ significantly from those in the Schwarzschild BH spacetime, with the most striking feature being that, in the spacetimes considered here, all tidal forces, both radial and angular, remain finite throughout the entire manifold. However, it is worth emphasizing that the Bardeen-type BB exhibited markedly different behavior compared to the other two cases, where significant compression effects were observed in certain regions. These effects were not present in the other scenarios, as evidenced by the analysis of the infinitesimal displacement vector.

This result suggests that spacetimes with richer and more complex causal structures, such as the Bardeen-type BB, which features not only a possible event horizon but also a Cauchy horizon, can induce a phenomenon in which the usual radial stretching and angular compression responsible for the well-known spaghettification of bodies is reversed. In this case, the test body undergoes compression along the radial direction and stretching in the angular direction. This behavior shows that BB backgrounds are, in principle, distinguishable from one another when we consider the effects of their geometry on test bodies subject to their influence.

A natural continuation of this work would be to investigate tidal disruption in BB spacetimes, identifying the conditions under which an extended body is torn apart by tidal forces near the center. This could lead to a generalized Roche limit adapted to these geometries, accounting for their regular core and causal structure. Another possible extension is the inclusion of rotation, which can modify the strength and direction of tidal forces and shift the location of key features such as extrema and sign changes. These developments may offer further insight into the physical properties of BBs and support efforts to distinguish them from classical BHs.

\section*{Acknowledgments}
\hspace{0.5cm} The authors would like to thank Conselho Nacional de Desenvolvimento Cient\'{i}fico e Tecnol\'ogico (CNPq) and Funda\c c\~ao Cearense de Apoio ao Desenvolvimento Cient\'ifico e Tecnol\'ogico (FUNCAP) for the financial support. This work is also supported by the Spanish National Grant PID2020-117301GA-I00 funded by MICIU/AEI/10.13039/501100011033 (``ERDF A way of making Europe" and ``PGC Generaci\'on de Conocimiento"); the Q-CAYLE project, funded by the European Union-Next Generation UE/MICIU/Plan de Recuperacion, Transformacion y Resiliencia/Junta de Castilla y Leon (PRTRC17.11) and the Department of Education, Junta de Castilla y León and FEDER Funds (Spain), Ref. CLU-2023-1-05.  This paper is also dedicated to the memory of my mother, Eloísa Mota (T. M. Crispim).
\bibliographystyle{apsrev4-1}
\bibliography{ref.bib}

\end{document}